\documentclass[pre,twocolumn,floatfix,superscriptaddress]{revtex4}
\usepackage{graphicx,amsfonts,amssymb,amsmath, hyperref}
\usepackage[applemac]{inputenc}

\newif\ifhyper
\hypertrue
\ifhyper
\hypersetup{
  citecolor = {green},
  colorlinks = {true}, 
  urlcolor = {blue} 
}
\fi

\newlength{\ldag}
\begin{document}

\title{Temperature Expansions in the Square Shoulder Fluid II: Thermodynamics}

\author{O. Coquand} 
\email{oliver.coquand@dlr.de}
\affiliation{Institut f\"ur Materialphysik im Weltraum, Deutsches Zentrum f\"ur Luft- und Raumfahrt (DLR), 51170 K\"oln, Germany}

\author{M. Sperl} 
\email{matthias.sperl@dlr.de}
\affiliation{Institut f\"ur Materialphysik im Weltraum, Deutsches Zentrum f\"ur Luft- und Raumfahrt (DLR), 51170 K\"oln, Germany}
\affiliation{Institut f\"ur Theoretische Physik, Universit\"at zu K\"oln, 50937 K\"oln, Germany}


\begin{abstract}

	In a companion paper \cite{Coquand19a}, we derived analytical expressions for the structure factor of the square-shoulder potential in a perturbative way
	around the high- and low-temperature regimes.
	Here, various physical properties of these solutions are derived.
	In particular, we investigate the large wave number sector, and relate it to the contact values of the pair-correlation function.
	Then, thermoelastic properties of the square-shoulder fluids are discussed.

\end{abstract}

\maketitle

\section{Introduction}

	The square-shoulder potential is one of the simplest purely repulsive potentials with a hard core.
	It can be expressed as:
	\begin{equation}
	\label{eqSqSh}
		U(r)=\left\{\begin{split}
			& +\infty \ , \\
			& U_0     \ , \\
			& 0       \ ,
		\end{split}\right. \!\!\!\!\!\!\!\!\!\!\!\!\!\!\!\!\!\!\!\!\!\!\!\!\!\!\!\!\!\!\!\!\!\!\!\!\!\!\!\!\!\!\!\!\!\!\!\!\!\!\!\!
		\begin{split}
			& 0\leqslant r<R \\
			& R \leqslant r < \lambda R \\
			& \lambda R\leqslant r \ ,
		\end{split}
	\end{equation}
	where $R$ is the particle's diameter, and $\lambda>1$.

	Such a potential shares with the hard-sphere potential the properties of being short-range, fully repulsive, and it has two natural hard-sphere limits
	($U_0\ll k_BT$ and $U_0\gg k_BT$); but the presence of an additional length scale significantly enriches the properties of such a system
	compared to the simple hard-sphere one.
	Indeed, beyond its simplicity that allows easier theoretical computations, a number of studies have demonstrated that the square-shoulder potential, and its smoother counterparts
	--- the so-called core-softened potentials --- can be of real physical significance in various physical contexts
	\cite{Young77,Levesque77,Denton97,Bolhuis97,Rascon97,Lang99,Jagla99,Malescio03,Ryzhov03,Osterman07,Pauschenwein08,Fomin08,Oliveira08,Gribova09,Buldyrev09,Zhou09,Fornleitner10,
	Sperl10,Fomin11,Yuste11,Ziherl11,Vilaseca11,Kim13,Khanpour13,Hus13,Gnan14,Dotera14,Boles16,Haro16,Gabrielse17,Pattabhiraman17a,Pattabhiraman17b,Pattabhiraman17c,Doukas18,Coquand19a}.

	First, it possesses a rich crystalline phase diagram \cite{Osterman07,Pauschenwein08,Fornleitner10,Ziherl11,Boles16,Doukas18}, that presents isostructural phase transitions
	\cite{Denton97,Bolhuis97}, allows for various type of crystal phases with high sensitivity to the value of $\lambda$ \cite{Gabrielse17,Pattabhiraman17c}, and even more involved structure such
	as stripe phases \cite{Malescio03,Pattabhiraman17a}, or quasicrytals \cite{Dotera14,Pattabhiraman17b}.
	Such behaviors are not purely theoretical and can be observed in a number of real systems such as Cesium or Cerium \cite{Young77}, nanocrystals \cite{Boles16}, or colloids \cite{Osterman07} for example.
	Some unusual behaviors have also been reported in the amorphous solid phase \cite{Fomin08,Sperl10}.

	Second, the fluid phase of repulsive potentials with two length scales is known to reproduce some water-like anomalies
	\cite{Jagla99,Ryzhov03,Oliveira08,Buldyrev09,Gribova09,Zhou09,Fomin11,Vilaseca11}, such as negative thermal expansion coefficient,
	re-entrant melting at high pressures or anomalous specific heat upon cooling (see \cite{Buldyrev09} and references therein).
	While the need for an attractive part in the potential to reproduce a genuine liquid-liquid phase transition is still debated \cite{Ryzhov03,Zhou09}, it is thus fair to claim that
	the square-shoulder fluid is a simple toy-model that displays a number of exotic properties.

	Moreover, as explained in \cite{Doukas18}, the success of the square-shoulder potential is not only due to its simplicity, but also to the universality of some of the phenomena it describes:
	although such a discontinuous potential is expected to be a poor description of many natural phenomena at the microscopic scale, it seems to capture most of the structural properties of the soft repulsive potentials,
	as long as the density of the fluid is low enough to ensure that multiple overlaps are scarce (note that $\lambda$ plays a key role here).
	In particular, the presence of a hump in the free energy of the square-shoulder system can explain the richness of its crystalline phase diagram \cite{Doukas18}.

	In spite of all this, very little is known about structural properties of the square-shoulder fluid from a theoretical point of view.
	In a previous paper \cite{Coquand19a}, we derived temperature expansions for the square-shoulder structure factor $S(q)$ (respectively organized in powers of $\Gamma=U_0/k_BT$ at high temperatures,
	and $p=\exp(-\Gamma)$ at low temperatures).
	In this paper, we discuss some of the physical properties of the square-shoulder fluid that can be deduced form these expansions.
	First, we focus on the large-$q$ sector of the structure factor which is related to the discontinuities of the pair-correlation function $g(r)$, and thus to the
	equation of state (from the pressure pathway \cite{Hansen06}).
	In particular, we show that such discontinuities can be read directly from the smooth asymptotic form of $S(q)$, which may be relevant when one wants to extract the discontinuities
	from a set of numerical data for example, where the very sharp variations and decrease of $g(r)$ can make it a quite delicate task.
	Then, we discuss the properties of the equations of state that can be built from $S(q)$ in both regimes, and which give access to the thermoelastic properties of the fluid.
	The fact that we can write them analytically allows us to assess their precision in the low-density regime by comparing them to the virial expansion.

	The paper is organized as follows: the first part discusses the large-$q$ asymptotic behavior of the structure factor.
	The second part discusses the equations of state.
	Finally we conclude.

\section{Large-$q$ behavior}

	The large-$q$ sector of the structure factor contains important information about the physics of the square-shoulder fluid.
	First, it is related to the sharpest variations of $g(r)$, such as its discontinuities, which are sufficient to
	derive the pressure equation of state of the fluid \cite{Smith77,Lang99}:
	\begin{equation}
	\label{eqPv}
		\frac{\beta P^v}{\rho}=1+\frac{2\pi R^3\rho}{3}\Big[g(R^+)+\lambda^3\big(g(\lambda R^+)-g(\lambda R^-)\big)\Big]\ .
	\end{equation}
	Second, there is evidence that the large-$q$ behavior of the structure factor could be related to the dynamics of arrest,
	that is the liquid-glass and glass-glass transitions in such systems \cite{Sperl10}.

	In this section, we decipher part of the information contained in the large-$q$ sector of the structure factor of the
	square-shoulder fluid.

	\subsection{Toy Model}

		First, recall that in the hard-sphere case, the large-$q$ behavior is given by:
		\begin{equation}
		\label{eqgq}                        
			S(q)\underset{q\rightarrow+\infty}{\sim}1+\rho\,\frac{4\pi\,R}{q^2}\,B(\varphi)\,\cos\left(q\,R\right)\ ,
		\end{equation}
		where $B(\varphi)=g(R^+)$ is the contact value, given by:
		\begin{equation}
		\label{eqCV}
			B(\varphi)=\frac{2+\varphi}{2(1-\varphi)^2}
		\end{equation}
		within the Percus-Yevick approximation.

\begin{figure}
			\begin{center}
				\includegraphics[scale=0.5]{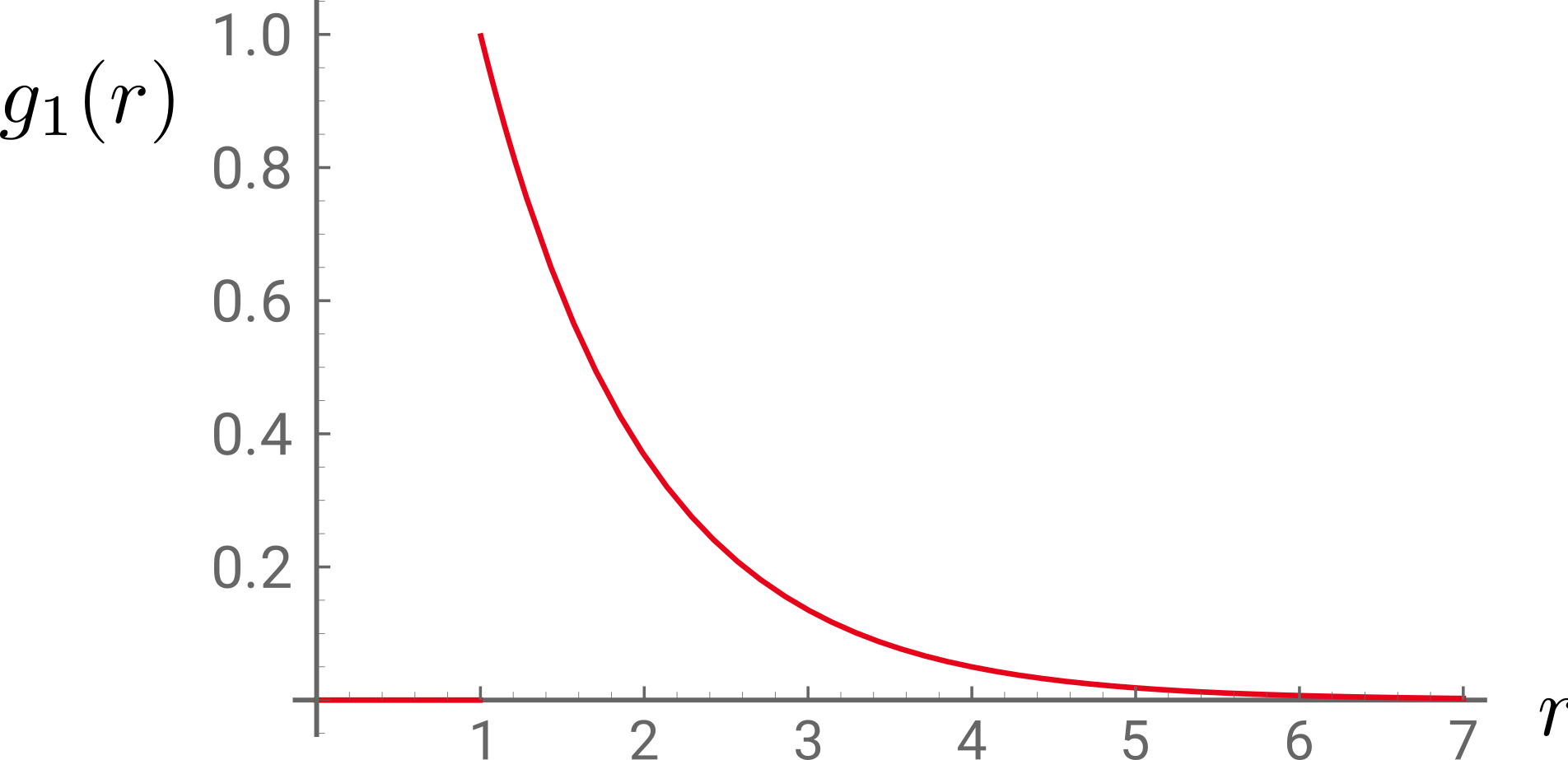}
			\end{center}
			\caption{Typical shape of the function $g_1$ given in Eq.~(\ref{eqg1}), represented here for $r_0=1$, $\Delta_g=1$, $l=1$.}
		\label{Figg1}
		\end{figure}

		In order to investigate what type of behavior is to be expected for the square-shoulder potential, we propose to study the following toy-model.
		Suppose that the pair-correlation function $g(r)$ of the system under investigation can be split as follows:
		\begin{equation}
			g(r)=g_0(r)+g_1(r)\ ,
		\end{equation}
		where $g_0$ is the pair correlation of a reference system, which is supposed to be known, and $g_1$ has the following form:
		\begin{equation}
		\label{eqg1}
			g_1(r)=\Delta_g\,\theta(r_0-r)\,e^{-(r-r_0)/l}\ ,
		\end{equation}
		the function $\theta$ being the Heaviside step function.
		The unknown added part consists of a discontinuous jump at $r=r_0$, whose value is determined by $\Delta_g$.
		The jump value is then exponentially damped, with a characteristic length $l$.
		The typical shape of such function is represented on Fig.~\ref{Figg1}.
		Note that we do not suppose that $g_1$ is a small perturbation term, the only hypothesis we make is that $\Delta_g$ is compatible
		with the constraint that $g(r)\geqslant0$. \\

		The corresponding structure factor is thus,
		\begin{equation}
			S(q)=1+\rho\int d^3r \big(g(r)-1\big)e^{i\vec{q}.\vec{r}}=S_0(q)+\Delta S_1(q)\ ,
		\end{equation}
		where $S_0(q)$ is the structure factor of the reference system.
		The remainder is then given by:
		\begin{equation}
			\begin{split}
				\Delta S_1(q)&=\rho\int d^3r\,g_1(r)\,e^{i\vec{q}.\vec{r}} \\
				&= \frac{4\pi\rho}{q^3} \int_0^{+\infty}du\, g_1\!\!\left(\frac{u}{q}\right)u\sin(u)\ .
			\end{split}
		\end{equation}
		The ansatz Eq.~(\ref{eqg1}) can then be used to get an explicit expression for the integral:
		\begin{equation}
			\Delta S_1(q)=\frac{4\pi\rho}{q^3}\,\Delta_g\,\int_{u_0}^{+\infty}du\,e^{-(u-u_0)/l_u}\,u\sin(u)\ ,
		\end{equation}
		where we have defined the dimensionless quantities $u_0=q\,r_0$ and $l_u=q\,l$.
		Finally, the following formula,
		\begin{equation}
			\begin{split}
				\int_a^be^{\alpha u}u\sin(u)&\,du=\Bigg[\bigg\{\bigg(\frac{2\alpha}{(1+\alpha^2)^2}-\frac{1}{1+\alpha^2}u\cos(u)\bigg) \\
				&+\bigg(\frac{1-\alpha^2}{(1+\alpha^2)^2}+\frac{\alpha}{1+\alpha^2}u\bigg)\sin(u)\bigg\}e^{\alpha u}\Bigg]_a^b,
			\end{split}
		\end{equation}
		with the replacement $\alpha\mapsto-1/l_u$ can be applied to yield the final result:
		\begin{equation}
			\begin{split}
				\Delta S_1(q)=&\frac{4\pi\rho}{q^3}\,\Delta_g\Bigg[\bigg(\frac{l_u^2}{l_u^2+1}\,u_0+\frac{2l_u^3}{(l_u^2+1)^2}\bigg)\cos(u_0) \\
				&+\bigg(\frac{l_u}{1+l_u^2}\,u_0+\frac{l_u^2(1-l_u^2)}{(1+l_u^2)^2}\bigg)\sin(u_0)\Bigg]\ ,
			\end{split}
		\end{equation}
		corresponding to the following large-$q$ behavior:
		\begin{equation}
			\Delta S_1(q)\underset{q\rightarrow+\infty}{\sim}\rho\frac{4\pi\,r_0}{q^2}\Delta_g\,\cos(q\,r_0)
		\end{equation}
		to be compared with Eq.~(\ref{eqgq}).

		All in all, the addition of an evanescent $\theta$-type discontinuity to the pair correlation function generates a large-$q$ leading
		contribution with a form very similar to the hard-sphere one, i.e. $\cos(q)/q^2$.
		The amplitude of the oscillations is determined both by the jump value $\Delta_g=g(r_0^+)-g(r_0^-)$, and the characteristic length
		at which the jumps shows up $r_0$, which also fixes the frequency of the oscillations.
		The result is independent of the damping length $l$, which indicates that the precise form of the damping function is of little importance
		in our reasoning, which may thus be generalized to any kind of sufficiently fast decreasing damping function.

	\subsection{Case of the Temperature Expansions}

		Let us begin with the low-temperature case.
		From the definition of $Q$:
		\begin{equation}
			S(q) = \big[Q(q)Q(-q)\big]^{-1}\,,
		\end{equation}
		and its explicit temperature expansion (Eq.~(19-22) of \cite{Coquand19a})
		the large-$q$ structure of $S$ can be worked out.
		As can be anticipated from the asymptotic behavior in the hard-sphere case Eq.~(\ref{eqgq}), the interesting behavior is captured by the Fourier transform
		of the direct correlation function $c_q$, related to $S(q)$ by:
		\begin{equation}
			S(q)=\frac{1}{1-\rho\,c_q}\ .
		\end{equation}
		In the low-temperature regime, and at first order in the $p$-expansion, its leading behavior can be written as:
		\begin{equation}
		\label{eqcqLT}
			c_q\underset{q\rightarrow+\infty}{\sim} A\,\frac{4\pi\,R}{q^2}+B_1\,\frac{4\pi\,R}{q^2}\cos(q\,R)+B_2\,\frac{4\pi\,d}{q^2}\,\cos(q\,d)\ ,
		\end{equation}
		where $d=\lambda R$.

		The coefficients of this decomposition are:
		\begin{equation}
			A=p\,\frac{3\phi\big(4\phi(\lambda-1)+1-\lambda^4\big)}{2\lambda^3(1-\phi)^2}\,,
		\end{equation}
		\begin{equation}
			B_1=p\,,
		\end{equation}
		\begin{equation}
			\begin{split}
				&B_2= \frac{2+\phi}{2(1-\phi)}\\
				&\;-p\,\frac{\big(6\phi(\phi-1)\lambda^2+\lambda(\lambda^3-6\phi)(2+7\phi)+3\phi(1+8\phi)\big)}{2\lambda^3(1-\phi)^2}\,,
			\end{split}
		\end{equation}
		where $\phi = \lambda^3 \varphi$ is the packing fraction of the outer core.
		It is a particularly relevant quantity in the low-temperature regime where it gives the packing fraction of the corresponding asymptotic hard-sphere system when $T\rightarrow0$.
		Note that a similar behavior was already pointed out in the case of a square-well potential \cite{Dawson00}, which is quite similar to the square-shoulder from this perspective.
		In the $p\rightarrow0$ limit, only a part of $B_2$ survives, and the expected hard-sphere result is recovered.

		From the study performed with help of our toy-model, it is natural to wonder how these coefficients are related to the discontinuities of the pair correlation function.
		The values of $g(r)$ in the vicinity of the discontinuities can be obtained directly from the Ornstein-Zernike equation since $Q$ is now explicit.
		The results are then as follows:
		\begin{equation}
			g(R^+)=p\,,
		\end{equation}
		\begin{widetext}
		\begin{equation}
			g(d^-)=p\frac{\big(6\lambda^2\phi(\phi-1)+4\lambda\phi(2+\phi)-3\phi(1+2\phi)+\lambda^4(2-3\phi-2\phi^2)\big)}{2\lambda^4(1-\phi)^2}\,,
		\end{equation}
		\begin{equation}
			g(d^+)=\frac{2+\phi}{2(1-\phi)^2}-p\frac{\phi\big(3+15\phi+\lambda^4(5+\phi)-8\lambda(1+2\phi)\big)}{\lambda^4(1-\phi)^2}\,.
		\end{equation}
		\end{widetext}
		The first equation yields a quantitatively wrong result, but should not come as a surprise as it is a direct consequence of the hypotheses we made
		to perform the low-temperature expansion.
		In the third one, the hard-sphere contact value appears explicitly, which ensures that everything is consistent in the low-temperature limit.
		Given the constraints $0\leqslant\phi\leqslant1$ and $\lambda\geqslant1$, the first order correction to $g(d^+)$ is always negative.

		Guided by the toy-model study, one can check that $B_1=g(R^+)$, and $B_2=g(d^+)-g(d^-)$, namely, the behavior of the square-shoulder system in the low-temperature
		limit is consistent with the hypothesis that the addition of a shoulder inside the core of a hard-sphere system modifies the pair correlation function in such
		a way that the main effects can be captured by the addition of a discontinuity at the inner-core diameter, the amplitude of which then quickly decreases.

		Moreover, the Eq.~(\ref{eqcqLT}) shows that the asymptotic oscillatory behavior of $c_q$ (or equivalently the structure factor) presents some beating
		phenomenon, with two frequencies fixed by the two length scales $R$ and $d$, and amplitudes depending on the jump values of the associated pair correlation function.
		This confirms the conjecture in \cite{Sperl10} that the contact values can be read from the characteristics of the large-$q$ beating signal.

		Finally, the pressure equation of state can be deduced from Eq.~(\ref{eqPv}) and the above results:
		\begin{widetext}
		\begin{equation}
		\label{eqPvLT}
			\frac{\beta P^v_{LT}}{\rho}=\frac{1+2\phi+3\phi^2}{(1-\phi)^2}-p\,\frac{2(\lambda-1)\phi\big(\lambda^3(\lambda+1)(2+7\phi)-3\phi(1+8\phi)+\lambda(2+\phi+6\phi^2)\big)}{\lambda^4(1-\phi)^2}\,.
		\end{equation}
		\end{widetext}
		The first term is the usual hard-sphere result in the Percus-Yevick approximation.
		The first order correction at low temperatures yields a small decrease in the fluid's pressure as one could have expected:
		when a shoulder is created inside the hard core, the fluid's particles tend to become softer, hence the reduced pressure.

		A similar study can be conducted on the high-temperature structure factor.
		The direct correlation function has the same asymptotic behavior as in Eq.~(\ref{eqcqLT}).
		For completeness, the full expression of the coefficients $A$, $B_1$ and $B_2$ are given in Appendix \ref{AqHT}.

		For sake of simplicity, we will not write here the full expressions of the values of $g$ around the discontinuities.
		However, it can be checked by using the form of $Q$ and the Ornstein-Zernike equation that $B_1=g(R^+)$ and $B_2=g(d^+)-g(d^-)$ still hold.
		These equations can also be used to derive the pressure equation of state in the high-temperature limit.
		The first order temperature correction to the hard-sphere limit yields a small increase in pressure, consistent with the intuition one can have
		about adding a repulsive shoulder to a hard-sphere system.

		Note that these results hold independent of the sign of $\Gamma$.
		As an aside result, we thus proved that a similar mechanism is at play in systems interacting via a square-well type of potential --- the attractive counterpart
		of the square-shoulder potential --- at least in the high-temperature limit which is common to both square-well and square shoulder systems.
		Indeed, the asymptotic form of the direct correlation function has been shown to be the same as Eq.~(\ref{eqcqLT}) (see Eq.~(12) of \cite{Dawson00}, or Eq.~(4) of \cite{Zaccarelli06}).
		Our results suggest that amplitude of the beatings observed in the square-well systems \cite{Dawson00,Zaccarelli03,Zaccarelli06} are directly related to the jump values of the pair correlation function
		in those systems as well (square-well systems have two jumps as well).
		Therefore, this type of large-$q$ behavior seems to be independent of the nature of the interaction --- attractive or repulsive --- and to be generic, to some extent,
		for interaction potentials with two length scales.

	\subsection{Discussion}

		From all the results above, and our toy-model, we can deduce that the main effect of the addition of finite-potential shoulder to a hard-sphere system -- be it
		inside or outside the hard-core -- on the pair correlation function is the addition of a discontinuity, whose amplitude gets strongly damped at large distances.

		However, it should be noted that:
		\begin{itemize}
			\item This explanation is not complete.
			Indeed, if the toy model reproduces well the observed behavior for the oscillating part of the large-$q$ tail, it does not explain the generation
			of the non-oscillating part characterized by $A$.

			\item It should not be understood the ansatz Eq.~(\ref{eqg1}) provides a complete description of the physics at play; it just captures the main effects.
			For example, let us apply the results of our toy-model to the hard-sphere fluid.
			We choose as reference system the zero-density limit of the system, whose pair correlation function is simply given by:
			\begin{equation}
				g_0^\infty(r)=\theta(r-R)\,.
			\end{equation}
			The corresponding structure factor is then
			\begin{equation}
			\label{eqSinfty}
				S_0^\infty(q)=1+\frac{4\pi\rho}{q^3}\big(qR\cos(qR)+\sin(qR)\big)\,.
			\end{equation}
			The finite density system is defined with an additional jump value of $\Delta_g=B(\varphi)-1$ located at $r=R$.
			From the previous study, we deduce that the correction to the asymptotic behavior is:
			\begin{equation}
				\Delta S_1(q)\underset{q\rightarrow+\infty}{\sim}\rho\frac{4\pi R}{q^2}\big(B(\varphi)-1\big)\cos(qR)\,,
			\end{equation}
			which combined with Eq.~(\ref{eqSinfty}) gives the well-known result.
			However, this does not mean that pair correlation function of a finite density hard-sphere fluid always have the form of Eq.~(\ref{eqg1}).
			In particular, additional oscillatory behavior is not captured by this ansatz.
			The form Eq.~(\ref{eqg1}) gives insight into the general form of the pair correlation function (jump at contact and relaxation towards 1)
			which drives the most prominent effects, but misses sub-leading corrections.
			Note that this general shape is precisely the information we kept to build our truncated pair correlation function (see Eq.~(28) of \cite{Coquand19a}).
			
		\end{itemize}

\section{Equations of state}

	We now discuss the thermodynamical properties of the structure factors computed in \cite{Coquand19a}.

	\subsection{Low-temperature expansion}

		\subsubsection{Compressibility equation of state}

			The compressibility equation of state is derived from the following equation \cite{Hansen06}:
			\begin{equation}
			\label{eqChiEOS}
				S(0)=\rho k_BT\,\chi_T=\frac{\chi_T}{\chi_0}\ ,
			\end{equation}
			where $\chi_T$ is the fluid's isothermal compressibility, and $\chi_0$ is the isothermal compressibility of an ideal gas.
			From this equation, one can derive that:
			\begin{equation}
			\label{eqPc}
				\frac{\beta P^c}{\rho}=\int_0^\varphi \frac{1}{S(0,\psi)}\,d\psi\ ,
			\end{equation}
			where it has been made explicit that the structure factor depends on the packing fraction of the system.
			Note that whenever the Percus-Yevick approximation is made, one should expect that the compressibility equation of state Eq.~(\ref{eqPc}),
			and the pressure equation of state Eq.~(\ref{eqPv}) yield different results \cite{Hansen06}.

			In the low temperature case, our previous expression of the structure factor can be integrated, to finally yield:
			\begin{widetext}
			\begin{equation}
			\label{eqPcLT}
				\begin{split}
					\frac{\beta P^c_{LT}}{\rho}=&\frac{1+\phi+\phi^2}{(1-\phi)^3}
					-p\frac{4(5\lambda^4-32\lambda+27)\ln(1-\phi)}{\phi \lambda^4} \\
					&+p\frac{2\big(\lambda^4(-10+23\phi-22\phi^2)+2\lambda(32-79\phi+65\phi^2)-27(2-5\phi+4\phi^2)\big)}{(1-\phi)^3 \lambda^4}\,.
				\end{split}
			\end{equation}
			\end{widetext}

			This is indeed different form Eq.~(\ref{eqPvLT}).
			The first term in Eq.~(\ref{eqPcLT}) is the usual Percus-Yevick result for hard spheres, the next terms are the first order corrections in $p$.

			Most importantly, knowing both equations of state enables us to build a third one whose natural $p\rightarrow0$ limit is the more accurate Carnahan-Starling
			equation of state, known to reproduce the hard-sphere equation of state \cite{Carnahan69} with a quite high accuracy.
			In order to do so, we must simply define \cite{Hansen06}:
			\begin{equation}
			\label{eqPCSLT}
				\frac{\beta P^{CS}_{LT}}{\rho} = \frac{2}{3} \frac{\beta P^{c}_{LT}}{\rho} + \frac{1}{3} \frac{\beta P^{v}_{LT}}{\rho}\ .
			\end{equation}
			It is thus expected that such an equation reproduces the physics of the square-shoulder fluid at low temperatures with a good accuracy.
			Interestingly, it is completely analytical.

\begin{widetext}

			\begin{figure}
				\begin{center}
					\includegraphics[scale=0.5]{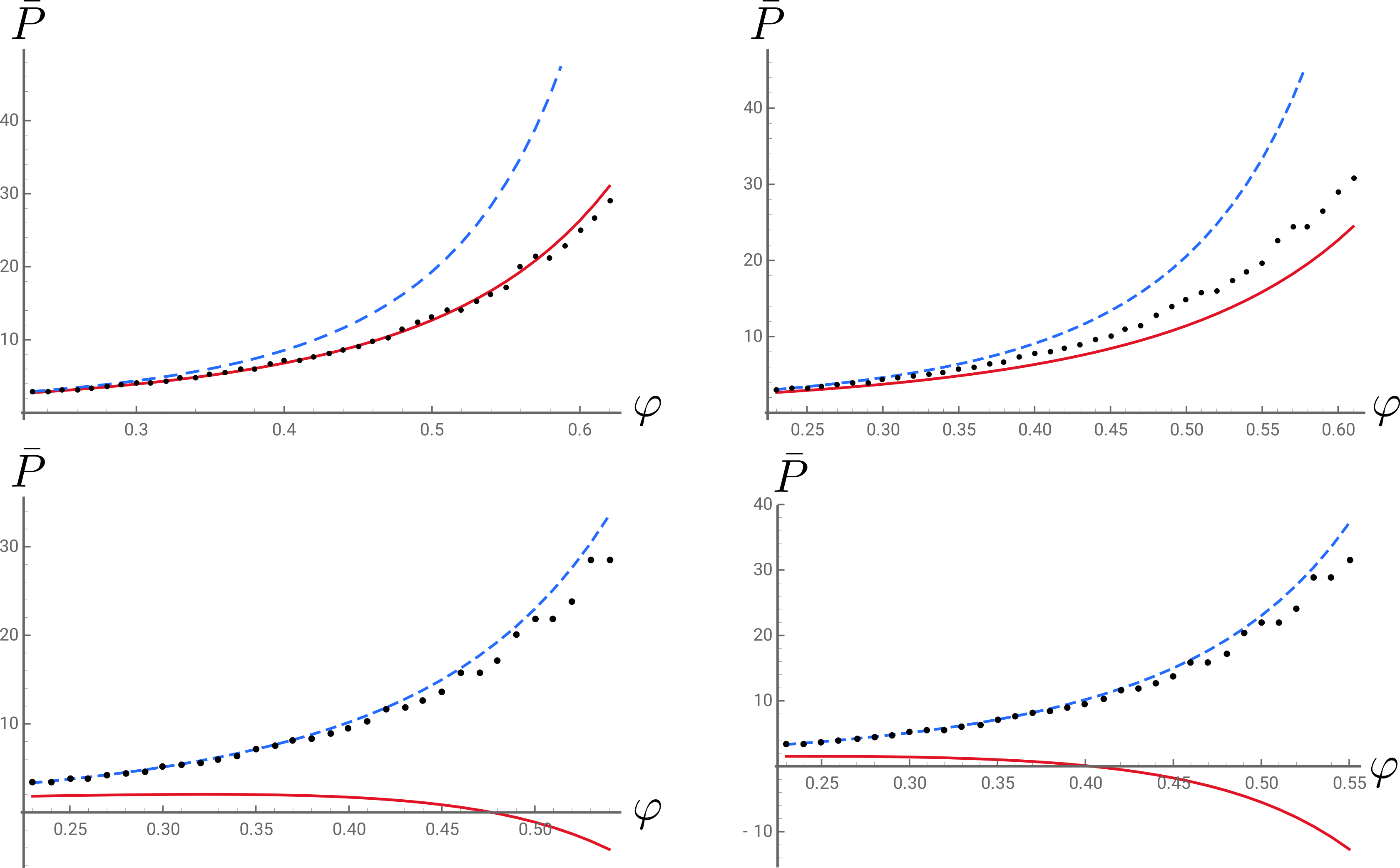}
				\end{center}
				\caption{Evolution of the dimesionless pressure $\bar{P}=\beta P/\rho$ for $\lambda=1.05$.
				The dashed blue line is the low-temperature result, the solid red line is the high-temperature result, the dots are data obtained
				within the Rogers-Young closure.
				Top-left: $\Gamma = 0.01$ ($T^*=100$); top-right: $\Gamma = 0.05$ ($T^* = 20$); bottom-left: $\Gamma = 4.5$ ($T^*\simeq2.2$); bottom-right: $\Gamma = 5.9$ ($T^* = 1.7$).}
				\label{figPd5}
			\end{figure}

			\end{widetext}

\begin{widetext}

			\begin{figure}
				\begin{center}
					\includegraphics[scale=0.5]{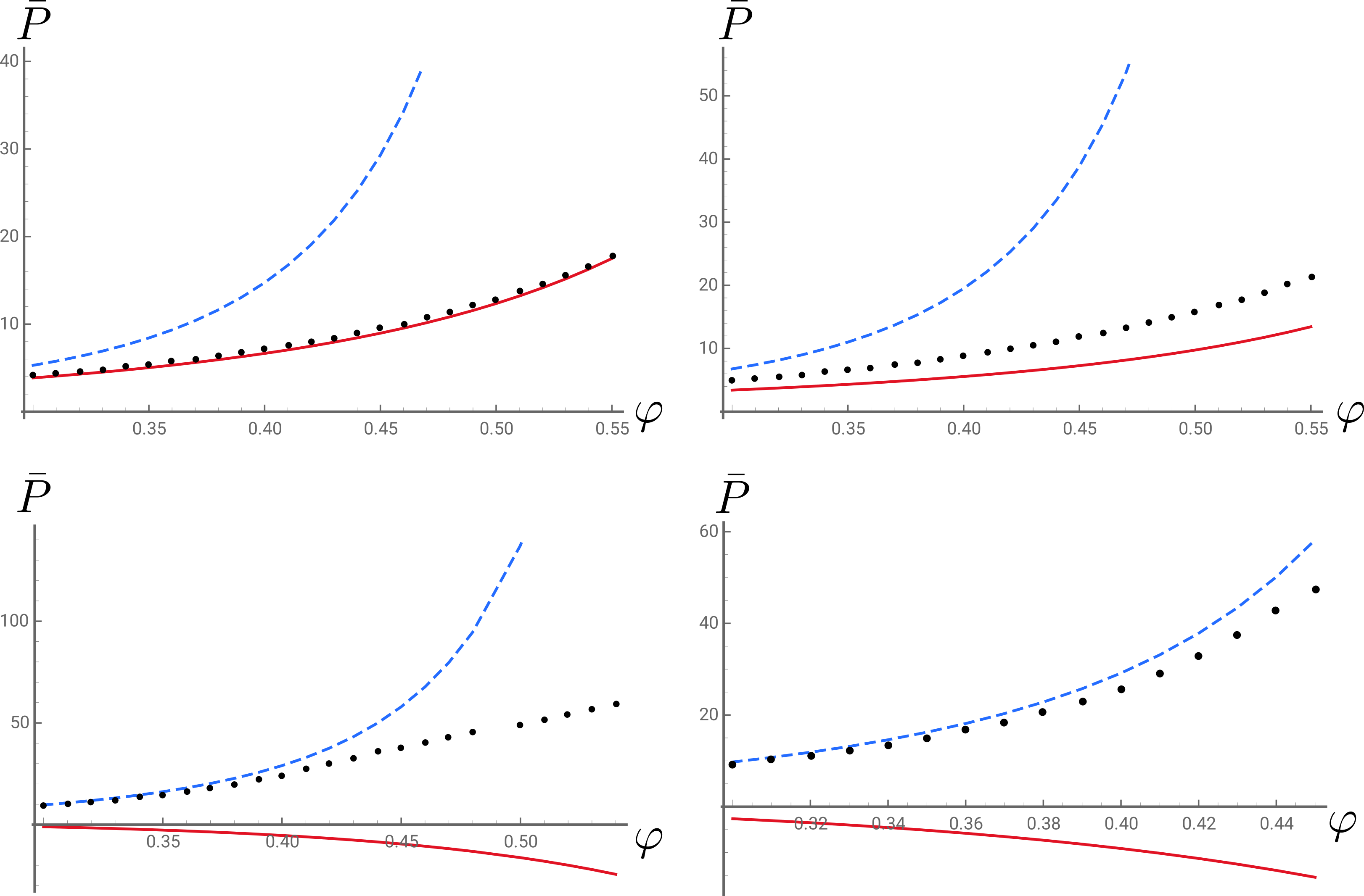}
				\end{center}
				\caption{Evolution of the dimesionless pressure $\bar{P}=\beta P/\rho$ for $\lambda=1.15$.
				The dashed blue line is the low-temperature result, the solid red line is the high-temperature result, the dots are data obtained
				within the Rogers-Young closure.
				Top-left: $\Gamma = 0.1$ ($T^*=10$); top-right: $\Gamma = 0.5$ ($T^* = 2$); bottom-left: $\Gamma = 4.5$ ($T^*\simeq2.2$); bottom-right: $\Gamma = 5.9$ ($T^* = 1.7$).}
				\label{figPd15}
			\end{figure}

			\end{widetext}

			In order to test the accuracy the equation of state Eq.~(\ref{eqPCSLT}), we compare it to data obtained by numerically solving the Ornstein-Zernike equation
			with the Roger-Young closure \cite{Rogers84}, which presents the advantage of being thermodynamically consistent.
			The results are displayed on Fig.~\ref{figPd5} and Fig.~\ref{figPd15}.

			For $\lambda$ very close to one, the agreement between the analytical Percus-Yevick and numerical Rogers-Young data appears to be quite good in
			appropriate regimes of temperature.
			However, when $\lambda$ increases, the agreement deteriorates as the packing fraction rises.
			A similar phenomenon was observed at the level of the structure factor \cite{Coquand19a}, it stems form the fact that for $\lambda$ and $\varphi$ big enough,
			$\phi$ grows so large that the low-temperature hard-sphere limit becomes ill-defined (see \cite{Coquand19a} for a more detailed discussion).

		\subsubsection{Low $\phi$ behavior}

			At low densities, the equation of state should match the virial expansion, which can be used to assess the precision of our approximation.
			Since we know that the hard-sphere part of the Carnahan-Starling equation of state only matches exactly the first three virial coefficients \cite{Carnahan69}, we will cut the $\phi$
			expansion at order $\phi^2$.
			In that case, Eq.~(\ref{eqPCSLT}) becomes:
			\begin{equation}
			\label{eqPCSLTV}
				\begin{split}
					\frac{\beta P^{CS}_{LT}}{\rho} \underset{\phi\ll1}{\simeq}&1 + 4\left(1+p\left(\frac{1}{\lambda^3}-1\right)\right)\phi \\
					&+ \left(10-2p\,\frac{109\lambda^4-18\lambda^2-136\lambda+45}{9\lambda^4}\right)\phi^2\,.
				\end{split}
			\end{equation}
			The virial expansion, on the other hand, yields:
			\begin{equation}
			\label{eqV}
				\frac{\beta P^V}{\rho} = 1+\mathcal{B}_2(T)\,\rho+\mathcal{B}_3(T)\,\rho^2\ ,
			\end{equation}
			where the coefficients $\mathcal{B}_i$ are known analytically for the square-shoulder potential at this order \cite{Kihara43}
			\footnote{In this last reference the computation is done for a square-well potential. However the expressions are exactly identical in case of a square-shoulder potential
			except for the sign of the potential in the region $[R;d]$.}
			Below, the expressions of the virial coefficients are given as a function of the coefficient $\gamma=1-e^{-\Gamma}$ which contains the temperature dependence.
			\begin{equation}
				\begin{split}
					& \mathcal{B}_2 = \frac{2\pi R^3}{3}\big(1+\gamma(\lambda^3-1)\big) \\
					& \mathcal{B}_3 = \frac{\pi^2R^6}{18}\big((5-15\gamma+16\gamma^2-6\gamma^3)+18\lambda^2\gamma^2(\gamma-1) \\
					&\ -32\lambda^3\gamma(\gamma-1)-18\lambda^4(1-\gamma)^2\gamma+\lambda^6\alpha(1-2\gamma+6\gamma^2)\big)
				\end{split}
			\end{equation}
			Then, taking into account the fact that $\phi=\lambda^3\varphi$, it appears that Eq.~(\ref{eqPCSLTV}) and Eq.~(\ref{eqV}) match perfectly at order $\phi$ if $p=1-\gamma$.
			Such a form of $p$ fulfills the requirements set by the low-temperature expansion.
			We will therefore from now on use this as explicit form of $p$.

			The expansion at order $\phi^2$ is not exactly equivalent to the virial expansion, even when $\mathcal{B}_3$ is expanded in powers of $p$:
			\begin{equation}
				\frac{\beta P^{CS}_{LT}}{\rho}-\frac{\beta P^V}{\rho}\simeq2p\,\frac{26\lambda^6+18\lambda^4-152\lambda^3+117\lambda^2-9}{9\lambda^6}\,.
			\end{equation}
			This should not come as a surprise: in order to be able not to make assumptions on $\phi$, some other parts of the structure factor have been approximated.

			In the case where the shoulder's width is small, the precision of our low-temperature truncation in the low density regime can be assessed as:
			\begin{equation}
				\frac{\beta P^{CS}_{LT}}{\rho}-\frac{\beta P^V}{\rho} = \frac{4\,p\,\delta}{3}+O(\phi^3,\delta^2,p^2)\,,
			\end{equation}
			where $\delta=\lambda-1$ characterizes the smallness of the shoulder.

		\subsubsection{Thermoelastic coefficients}

			From the equation of state Eq.~(\ref{eqPCSLT}) it is possible to compute the different thermoelastic coefficients, which fully characterize the elastic response
			of the fluid as a function of temperature in this regime.
			We recall that in this framework increasing the temperature is equivalent to inducing a softening of the outer core of our system of particles, which is completely hard at zero temperature.

			First, the isothermal compressiblity can be obtained from Eq.~(\ref{eqChiEOS}).
			In the following, in order to lighten notations, we will work with dimensionless quantities, denoted with a bar.
			Hence, the dimensionless isothermal compressiblity can be defined as $\bar{\chi}_T=\chi_T/\chi_0$.
			Its $p$-expansion reads:
			\begin{widetext}
			\begin{equation}
			\label{eqChiLT}
				\begin{split}
					\bar{\chi}_T=\frac{(1-\phi^4)}{\mathcal{P}_1(\phi)}-\frac{2p\,\phi(1-\phi)^4}{3\lambda^4\mathcal{P}_1(\phi)}\big[&\lambda^4(12\phi^4-49\phi^3+56\phi^2-61\phi-12)
					-6\lambda^2\phi(1-\phi)^2(2\phi-3) \\
					&+4\lambda(15\phi^4-44\phi^3+58\phi^2+22\phi+3)-3\phi(20\phi^3-61\phi^2+80\phi+15)\big]\,,
				\end{split}
			\end{equation}
			\end{widetext}
			where
			\begin{equation}
			\label{eqPphi}
				\mathcal{P}_1(\phi)=1+4\phi+4\phi^2-4\phi^3+\phi^4\,.
			\end{equation}

\begin{figure}
				\begin{center}
					\includegraphics[scale=0.5]{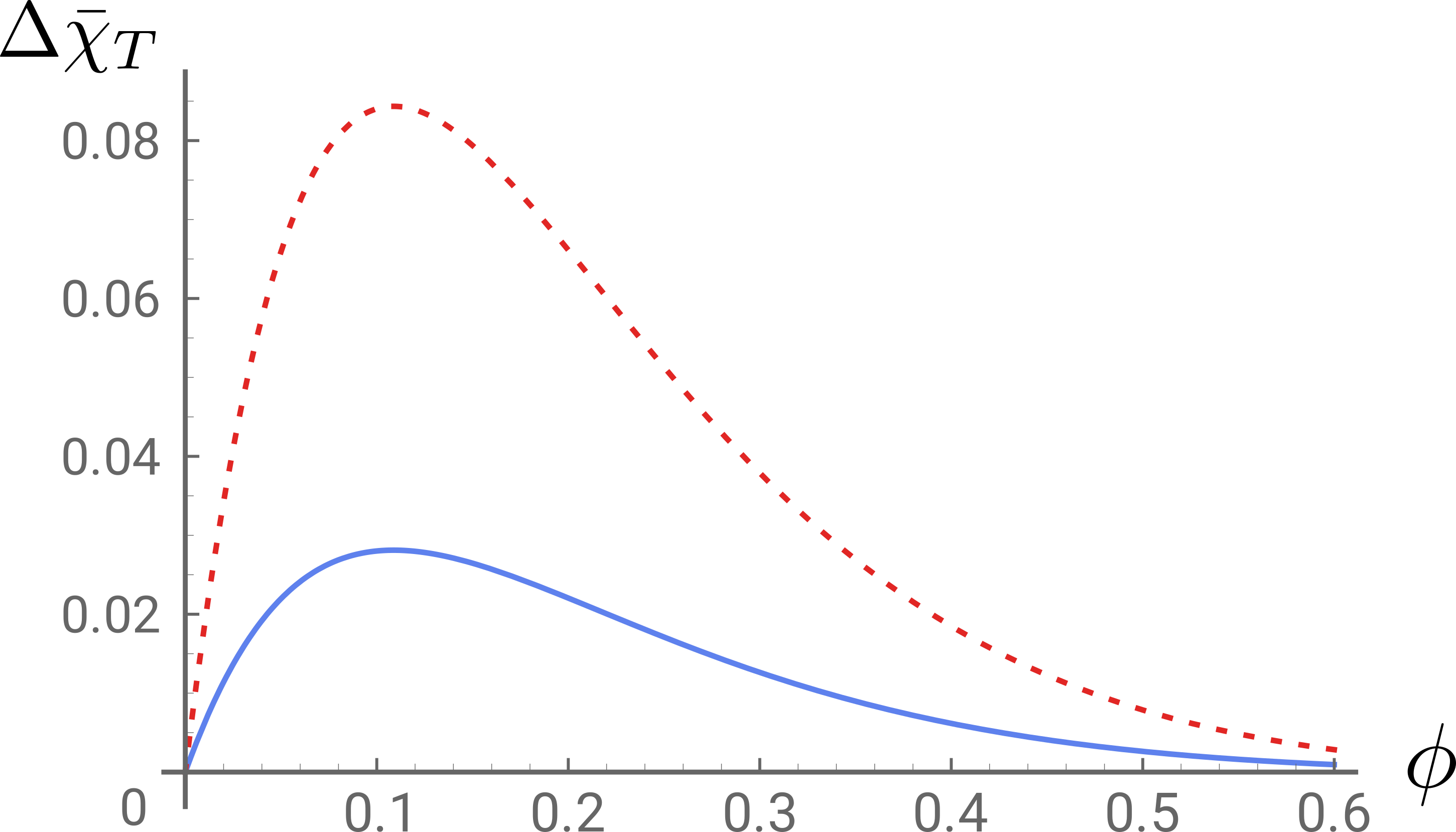}
				\end{center}
				\caption{Evolution of the dimensionless correction to the isothermal compressiblity with the outer-core packing fraction for $\lambda=1.2$.
				The full line corresponds to $p=0.2$ ($T^*=0.62$), and the dotted one to $p=0.6$ ($T^*=1.96$).}
				\label{FigDeltaChiT}
			\end{figure}

			Notice that the first term in Eq.~(\ref{eqChiLT}) is exactly the result expected for the Carnahan-Starling equation for hard-spheres \cite{Lee95}.
			We will therefore split the expression of $\bar{\chi}_T$ into the hard-sphere compressibility, and the low-temperature correction, that we shall denote $\Delta\bar{\chi}_T$.
			Its evolution is represented on Fig.~\ref{FigDeltaChiT}.
			The first order correction being positive means that softening the core of the particles leads to an increase in the fluid's compressiblity, in agreement with physical intuition.\\

			Next, the thermal expansion coefficient $\alpha$ can be computed.
			It is defined as:
			\begin{equation}
				\alpha=\frac{1}{V}\left(\frac{\partial V}{\partial T}\right)_P\,.
			\end{equation}
			To define its dimensionless counterpart, we first define a dimensionless temperature $T^*=1/\Gamma=k_BT/U_0$.
			The dimensionless thermal expansion coefficient is thus $\bar{\alpha}=\alpha\times k_B/U_0$.
			It reads:
			\begin{widetext}
			\begin{equation}
			\label{eqAlphaLT}
				\begin{split}
					\bar{\alpha}&= \frac{\mathcal{P}_2(\phi)}{T^*\mathcal{P}_1(\phi)}+\frac{2\,p}{3(T^*)^2\lambda^4\phi\mathcal{P}_1(\phi)^2}\bigg\{
					(\lambda-1)(\phi-1)\phi\Big[\mathcal{P}_1(\phi)\big(\lambda(20-44\phi+49\phi^2-7\phi^3) \\
					&\:+\lambda^2(\lambda+1)(20-44\phi+55\phi^2-19\phi^3+6\phi^4)+3(-36+90\phi-71\phi^2-11\phi^3+10\phi^4)\big)\\
					&+2T^*(1-\phi)\big(\lambda(10+22\phi-29\phi^2-86\phi^3+81\phi^4-31\phi^5) \\
					&\:+\lambda(\lambda+1)(10+22\phi-35\phi^2-71\phi^3+81\phi^4-37\phi^5-6\phi^6+3\phi^7) \\
					&\:+3(5\phi^7-10\phi^6+44\phi^5-153\phi^4+187\phi^3-35\phi^2-45\phi-18)\big)\Big] \\
					&-4(1+T^*)\mathcal{P}_3(\lambda)(1-\phi)^4\mathcal{P}_1(\phi)\ln(1-\phi)\bigg\}\,,
				\end{split}
			\end{equation}
			\end{widetext}
			where we used the following functions:
			\begin{equation}
			\label{eqPphi2}
				\begin{split}
					& \mathcal{P}_2(\phi) = 1-2\phi^3+\phi^4 \\
					& \mathcal{P}_3(\lambda) = 5\lambda^4-32\lambda+27\,.
				\end{split}
			\end{equation}

\begin{figure}
				\begin{center}
					\includegraphics[scale=0.5]{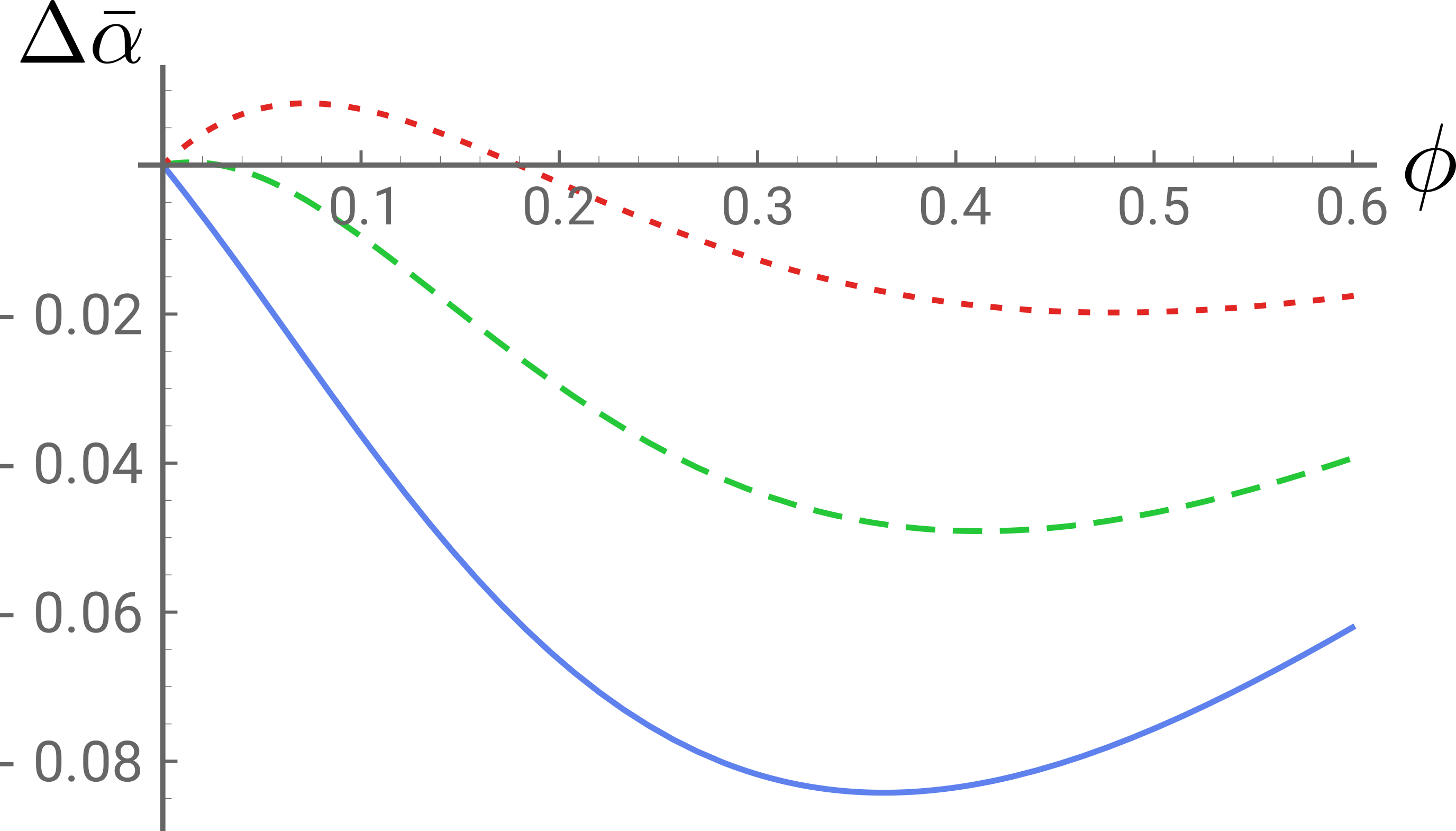}
				\end{center}
				\caption{Evolution of the dimensionless correction to the thermal expansion coefficient with the outer-core packing fraction for $\lambda=1.2$.
				The full line corresponds to $p=0.2$ ($T^*=0.62$), the dashed one to $p=0.4$ ($T^*=1.09$) and the dotted one to $p=0.6$ ($T^*=1.96$).}
				\label{FigAlpha}
			\end{figure}

			Once again, the first term corresponds to the Carnahan-Starling hard-sphere result \cite{Wilhelm74}.
			Hence, a $\Delta\bar{\alpha}$ containing the first corrections to hard spheres can be defined.
			Its evolution is plotted on Fig.~\ref{FigAlpha}.
			In most cases, the first order correction is negative, what means that softening the spheres reduces the fluid's ability to dilate upon temperature increases.
			Interestingly, for high enough values of $p$ (typically $p\gtrsim 0.4$, see Fig.~\ref{FigAlpha}) a region emerges at low density where this correction changes sign.
			This shows that something non-trivial occurs in the fluid at moderate temperatures.
			We shall discuss this a bit more later.

			Note that such a behavior is only possible if $\Delta\bar{\alpha}$ has a non-trivial $p$-dependence, what is not obvious at first glance from Eq.~(\ref{eqAlphaLT}).
			However, it should be kept in mind that $T^*=-1/\ln(p)$, so that in addition to the expansion in powers of $p$, the thermal expansion coefficient also includes
			subdominant logarithmic -- and square-logarithmic -- corrections, which in turn allow for a richer phenomenology as the temperature is varied. \\

\begin{figure}
				\begin{center}
					\includegraphics[scale=0.5]{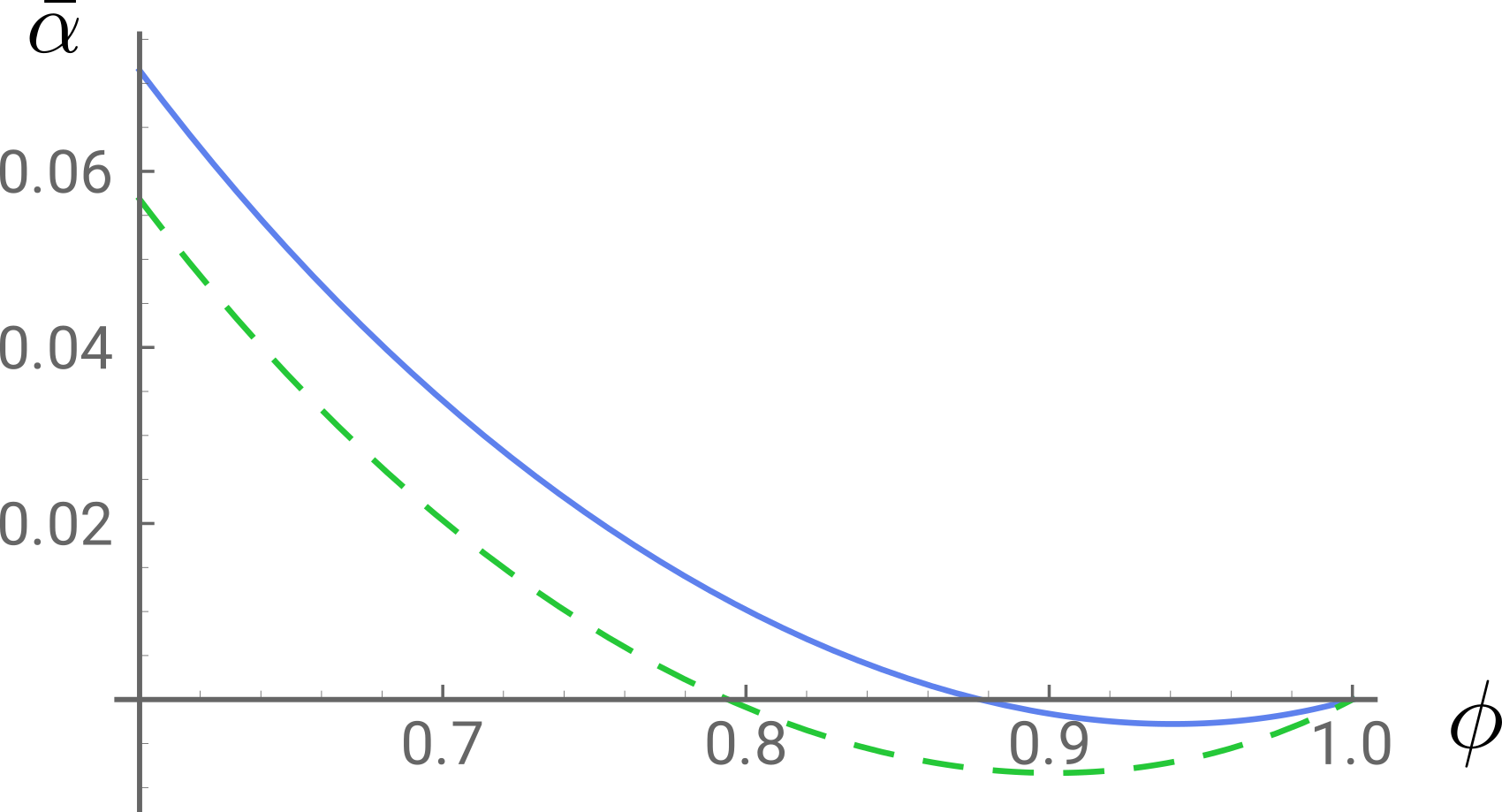}
				\end{center}
				\caption{Evolution of the dimensionless thermal expansion coefficient with the outer-core packing fraction for $\lambda=1.75$ (full line)
				and $\lambda=1.95$ (dashed line).
				The temperature is given by $p=0.3$ ($T^*=0.83$).}
				\label{FigAlphaLT}
			\end{figure}

\begin{figure}
				\begin{center}
					\includegraphics[scale=0.5]{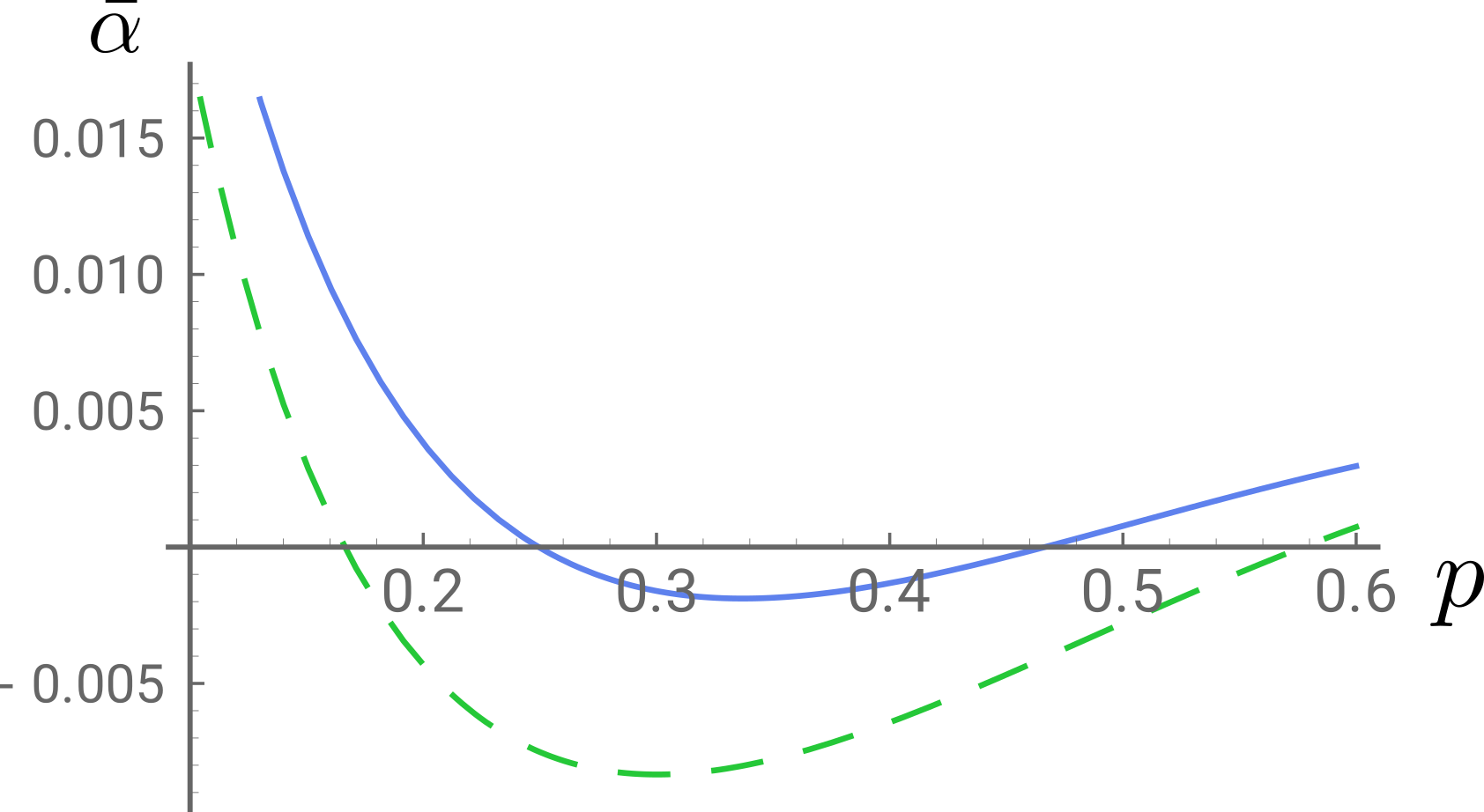}
				\end{center}
				\caption{Evolution of the dimensionless thermal expansion coefficient with $p$ for $\phi=0.9$.
				The full line corresponds to $\lambda=1.75$ ($\varphi\simeq0.17$) and the dashed line to $\lambda=1.95$ ($\varphi\simeq0.12$).}
				\label{FigAlphaLTp}
			\end{figure}

			The full thermal expansion coefficient is also an interesting quantity, since it is related to some anomalies of the square-shoulder fluid \cite{Buldyrev09}.
			In particular, we want to know whether it can be negative for some regimes of parameters.
			In the case of the low temperature expansion, such case can indeed occur, as shown on Fig.~\ref{FigAlphaLT} and Fig.~\ref{FigAlphaLTp}, for some non trivial
			range of packing fraction and temperature.
			Such a behavior however does not show up for values of $\lambda$ close to 1, hence the unusually high values chosen on the Figures.

			First, Fig.~\ref{FigAlphaLT} displays the evolution of $\bar\alpha$ with $\phi$.
			In that case, if $\lambda$ is big enough (typically bigger than 1.5), a region of negative $\bar\alpha$ appears at high packing fractions, that grows bigger and bigger
			as $\lambda$ is increased.
			One should not be confused by the very high values of $\phi$ involved, here the shoulder width is quite large, so that they typically correspond
			to quite low values of the packing fractions of the hard cores (see caption of Fig.~\ref{FigAlphaLTp} for example).
			It should also be noted that in our formalism, where the low-temperature expressions are built from their close hard sphere limit, $\phi$ cannot be bigger than 1,
			no matter how small $\varphi$ is.
			Anyway, it is expected that if this condition is not fulfilled, the structure of the fluid (or solid) is driven by many-body effects not captured by our simple
			potential \cite{Doukas18}.

			Then, the behavior of $\bar\alpha$ with respect to the temperature is examined on Fig.~\ref{FigAlphaLTp}.
			It appears that the negativity is never kept up to arbitrarily low temperature.
			This should not come as a surprise: it is clear from Eq.~(\ref{eqAlphaLT}) that the very-low temperature regime is controlled by the hard-sphere contribution which
			is always positive.
			The subleading term is of order $p\ln(p)^2$, which means that: (i) it actually vanishes as $p\rightarrow0$, but (ii) it can easily grow quite large as we go away from
			that limit.
			Because of such logarithmic corrections, it is expected that numerical values of $\bar\alpha$ are much more sensitive to higher orders than the other quantities we
			discussed so far.
			Some caution is thus required when interpreting the numbers showing up on those charts. \\

			Finally, we compute the isochoric compression coefficient $\beta_V$, defined as:
			\begin{equation}
			\label{eqBetadef}
				\beta_V=\frac{1}{P}\left(\frac{\partial P}{\partial T}\right)_V\,.
			\end{equation}
			As for $\alpha$, a dimensionless counterpart is defined through $\bar{\beta}_V=\beta_V\times k_B/U_0$.
			Its expression is then,

			\begin{widetext}
			\begin{equation}
			\label{eqBetaVLT}
				\begin{split}
					\bar{\beta}_V&=\frac{1}{T^*}-\frac{p}{3(T^*)^2\lambda^4\phi(1+\phi+\phi^2-\phi^3)}\Big\{2\phi\big[\lambda^4(20-44\phi+55\phi^2-19\phi^3+6\phi^4)-6\lambda^2\phi^2(1-\phi)^2 \\
					&+2\lambda(15\phi^4-13\phi^3-131\phi^2+157\phi-64)+3(36-90\phi+71\phi^2+11\phi^3-10\phi^4)\big]+8\mathcal{P}_3(\lambda)(1-\phi)^3\ln(1-\phi)\Big\}\,.
				\end{split}
			\end{equation}
			\end{widetext}

\begin{figure}
				\begin{center}
					\includegraphics[scale=0.5]{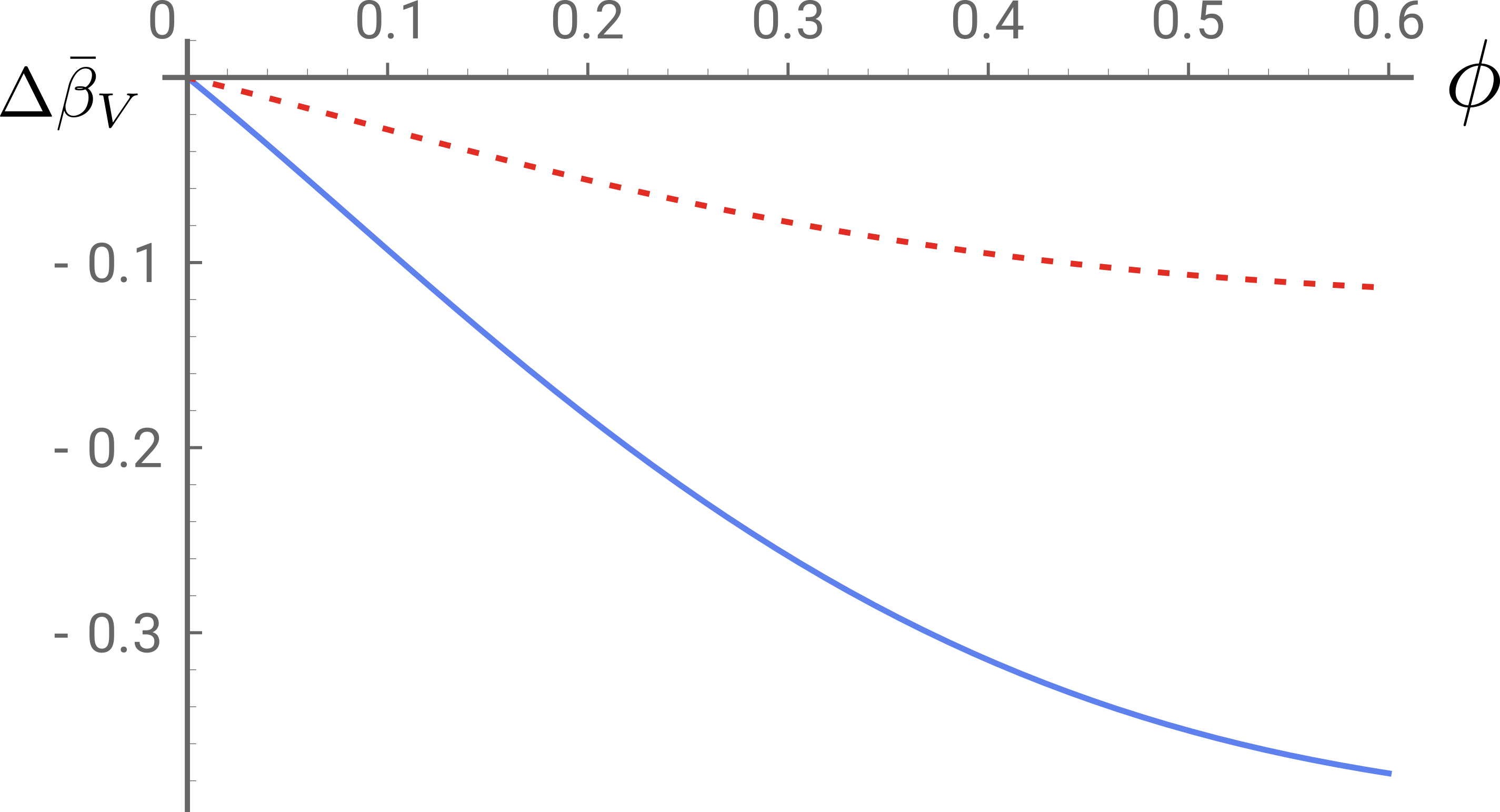}
				\end{center}
				\caption{Evolution of the dimensionless correction to the isochoric compression coefficient with the outer-core packing fraction for $\lambda=1.2$.
				The full line corresponds to $p=0.2$ ($T^*=0.62$), and the dotted one to $p=0.6$ ($T^*=1.96$).}
				\label{FigDeltaBetaV}
			\end{figure}

			In the Carnahan-Starling equation of state, the right-hand side depends on $\phi$ only \cite{Carnahan69}.
			Since in the definition Eq.~(\ref{eqBetadef}) $\beta_V$ is defined at constant volume, that is constant $\phi$, then the hard-sphere Carnahan-Starling value of $\beta_V$
			is simply $1/T$, in agreement with the first term of Eq.~(\ref{eqBetaVLT}).
			Once again, the explicit presence of $T^*$ gives rise to logarithmic corrections to the $p$-expansion.
			The correction to the hard-sphere result $\Delta\bar{\beta}_V$ is well-defined, and is represented on Fig.~\ref{FigDeltaBetaV}.
			It being always negative means that softening the particles leads a reduced ability to increase pressure as the temperature rises.
			With the kinetic pressure picture in mind, such a result seems reasonable: softer particles, when more agitated increase the inner-pressure of the fluid less efficiently as if they were hard.

			All these results can be put together to check for thermodynamical consistency.
			Indeed, from Eq.~(\ref{eqPCSLT}), Eq.~(\ref{eqChiLT}), Eq.~(\ref{eqAlphaLT}) and Eq.~(\ref{eqBetaVLT}), it can be checked that:
			\begin{equation}
			\label{eqConst}
				\alpha=P\,\beta_V\,\chi_T\,,
			\end{equation}
			as required.

			\begin{figure}
				\begin{center}
					\includegraphics[scale=0.5]{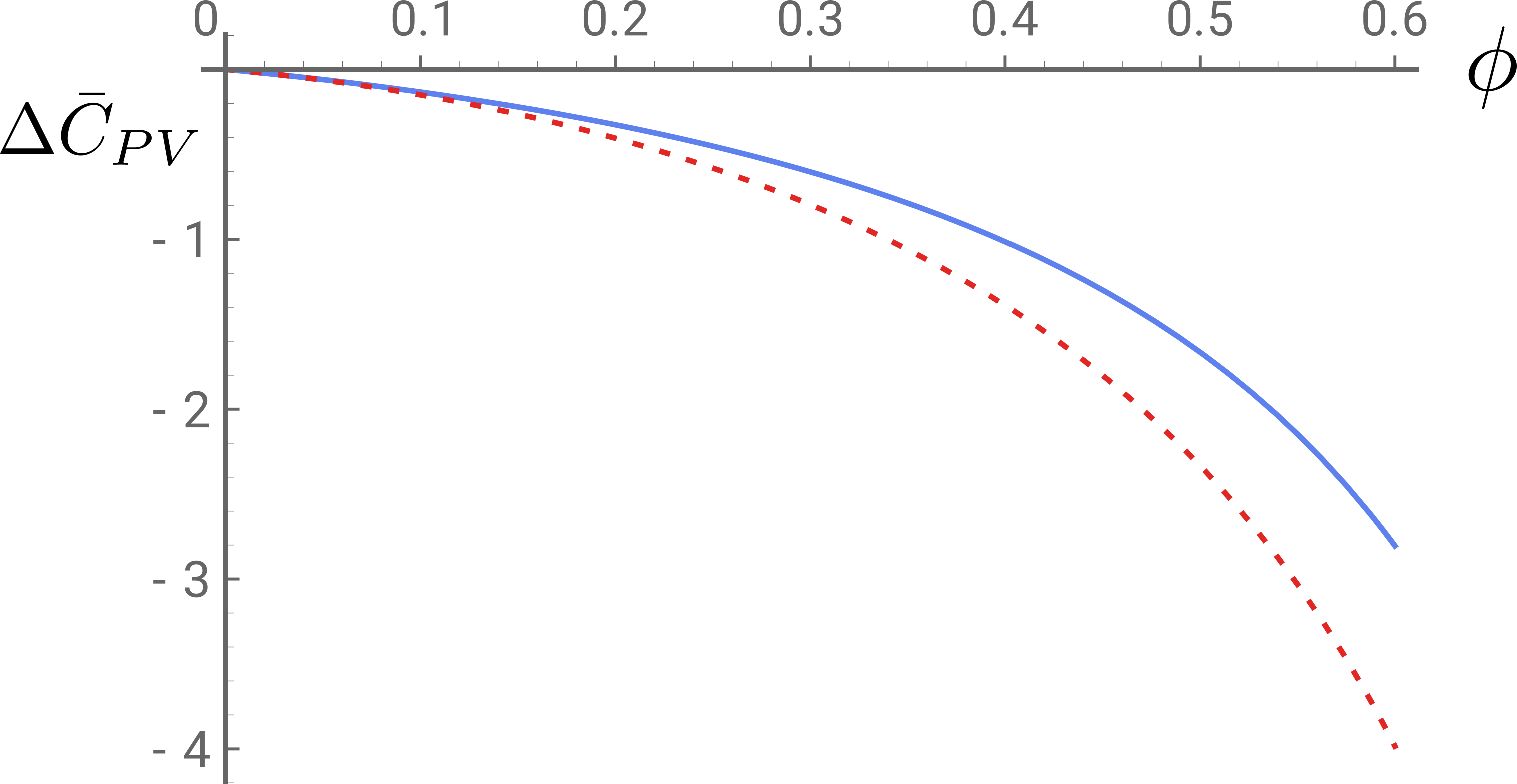}
				\end{center}
				\caption{Evolution of the dimensionless correction to the heat capacity difference with the outer-core packing fraction for $\lambda=1.2$.
				The full line corresponds to $p=0.2$ ($T^*=0.62$), and the dotted one to $p=0.6$ ($T^*=1.96$).}
				\label{FigDeltaCpCv}
			\end{figure}
			
			As a final result, we can combine all these equations with Mayer's relation to get the difference between the two molar heat capacities $C_P$ and $C_V$.
			The Eq.~(\ref{eqConst}) can be used to further simplify the relation:
			\begin{equation}
				\bar{C}_P-\bar{C}_V = (T^*)^2 \frac{\bar{\alpha}^2}{\bar{\chi}_T}\,,
			\end{equation}
			where we defined the dimensionless molar heat capacities by dividing by the ideal gas constant $R$.
			All in all, the heat capacities difference results in:

			\begin{widetext}
			\begin{equation}
			\label{eqCpCvLT}
				\begin{split}
					\bar{C}_P-\bar{C}_V &= \frac{\mathcal{P}_2(\phi)^2}{(1-\phi)^4\mathcal{P}_1(\phi)}+\frac{2\,p\,\mathcal{P}_2(\phi)}{3\lambda^4\phi(1-\phi)^4\mathcal{P}_1(\phi)^2}
					\Bigg\{\phi^2\mathcal{P}_2(\phi)\big(\lambda^4(12\phi^4-49\phi^3+56\phi^2-61\phi-12) \\
					&\: -6\lambda^2\phi(1-\phi)^2(2-3\phi)+4\lambda(3+22\phi+58\phi^2-41\phi^3+15\phi^4)-3\phi(15+80\phi-61\phi^2+20\phi^3)\big)\\
					&-\frac{2}{T^*}\bigg[\phi(1-\phi)(\lambda-1)\Big(\mathcal{P}_1(\phi)\big(\lambda(20-44\phi+49\phi^2-7\phi^3)+\lambda^2(\lambda+1)(20-44\phi+55\phi^2-19\phi^3+6\phi^4) \\
					&\:+3(10\phi^4-11\phi^3-71\phi^2+90\phi-36)\big)+2T^*(1-\phi)\big(\lambda(10+22\phi-29\phi^2-86\phi^3+81\phi^4-31\phi^5) \\
					&\:+\lambda^2(\lambda+1)(10+22\phi-35\phi^2-71\phi^3+81\phi^4-37\phi^5-6\phi^6+3\phi^7) \\
					&\:+3(5\phi^7-10\phi^6+44\phi^5-153\phi^4+187\phi^3+35\phi^2-45\phi-18)\big)\Big) 
					+4(1+T^*)\mathcal{P}_3(\lambda)(1-\phi)^4\mathcal{P}_1(\phi)\ln(1-\phi)\bigg]\Bigg\}\,.
				\end{split}
			\end{equation}
			\end{widetext}
			The first term of this equation is the expected result for Carnahan-Starling hard spheres \cite{Wilhelm74}.
			A first order correction due to the softness of the outer core can be therefore defined, it is denoted $\Delta\bar{C}_{PV}$.
			Its evolution with $\phi$ is represented on the Fig.~\ref{FigDeltaCpCv}.
			The first order correction is always negative, which is to be related to the lesser ability of the fluid to expand when $T$ increases.

		\subsubsection{Spinodal line}

			The introduction of a new length scale in the hard-sphere potential allows for a richer phase diagram.
			As a matter of fact, it is \textit{a priori} possible to define two fluid phases for the square-shoulder system: one with particle diameter $R$, and one with
			particle diameter $d$, called in the following low-density and high-density fluid respectively.
			While it is expected that at low temperatures the particles are typically too hard to allow for the existence of the low-density fluid, it is still possible to investigate
			whether our solution allows for such a transition at moderate temperature.
			Obviously, this requires to extrapolate the solution a bit beyond the regime in which it is expected to be accurate.
			Therefore, the following results should be treated with caution: the aim is more to establish a qualitative picture than to give reliable quantitative values.

			The prospective fluid-fluid transition will be investigated through the study of a possible spinodal line between the two phases.
			The latter is characterized by a diverging isothermal compressibility, or equivalently a vanishing $1/S(0)$ according to Eq.~(\ref{eqChiEOS}).
			For sake of simplicity, we will perform this study with use of this latter equation, so that our expression for $\chi_T$ is a bit different from Eq.~(\ref{eqChiLT}).

			The temperature at which the condition $1/S(0)=0$ is realized corresponds to $p=p_0$, with
			\begin{equation}
			\label{eqp0}
				p_0=\frac{\lambda^4(1+2\phi)}{2\phi(\lambda-1)\big(4\lambda(1\!+\lambda\!+\lambda^2)+\phi(5\lambda^3\!+5\lambda^2\!+\!5\lambda\!-\!27)\big)}\,.
			\end{equation}
			This solution must moreover be such that $0\leqslant p_0\leqslant 1$ (by definition of $p=\exp(-\Gamma)$).
			The first inequality is automatically satisfied by any $p_0$ obeying Eq.~(\ref{eqp0}).
			Thus only the second one can be violated if $\phi$ and $\lambda$ are not properly chosen.
			The limiting case condition $p_0=1$ is a simple polynomial of order 2 in $\phi$ whose roots are
			\begin{equation}
				\phi_{\pm}=\frac{\lambda(4-3\lambda^3)\pm\sqrt{\mathcal{P}_\Delta(\lambda)}}{2\mathcal{P}_2(\lambda)}\,,
			\end{equation}
			with
			\begin{equation}
				\mathcal{P}_\Delta(\lambda)=19\lambda^8-88\lambda^5+54\lambda^4+16\lambda^2\,.
			\end{equation}
			The roots $\phi_\pm$ are well defined only as long as $\mathcal{P}_\Delta(\lambda)\geqslant0$.
			This condition is always true except in the interval $[\lambda_1^*;\lambda^*_2]$, where $\lambda_1^*\simeq1.03$ and $\lambda^*_2\simeq1.23$ are the two real roots of $\mathcal{P}_\Delta$.

\begin{figure}
				\begin{center}
					\includegraphics[scale=0.5]{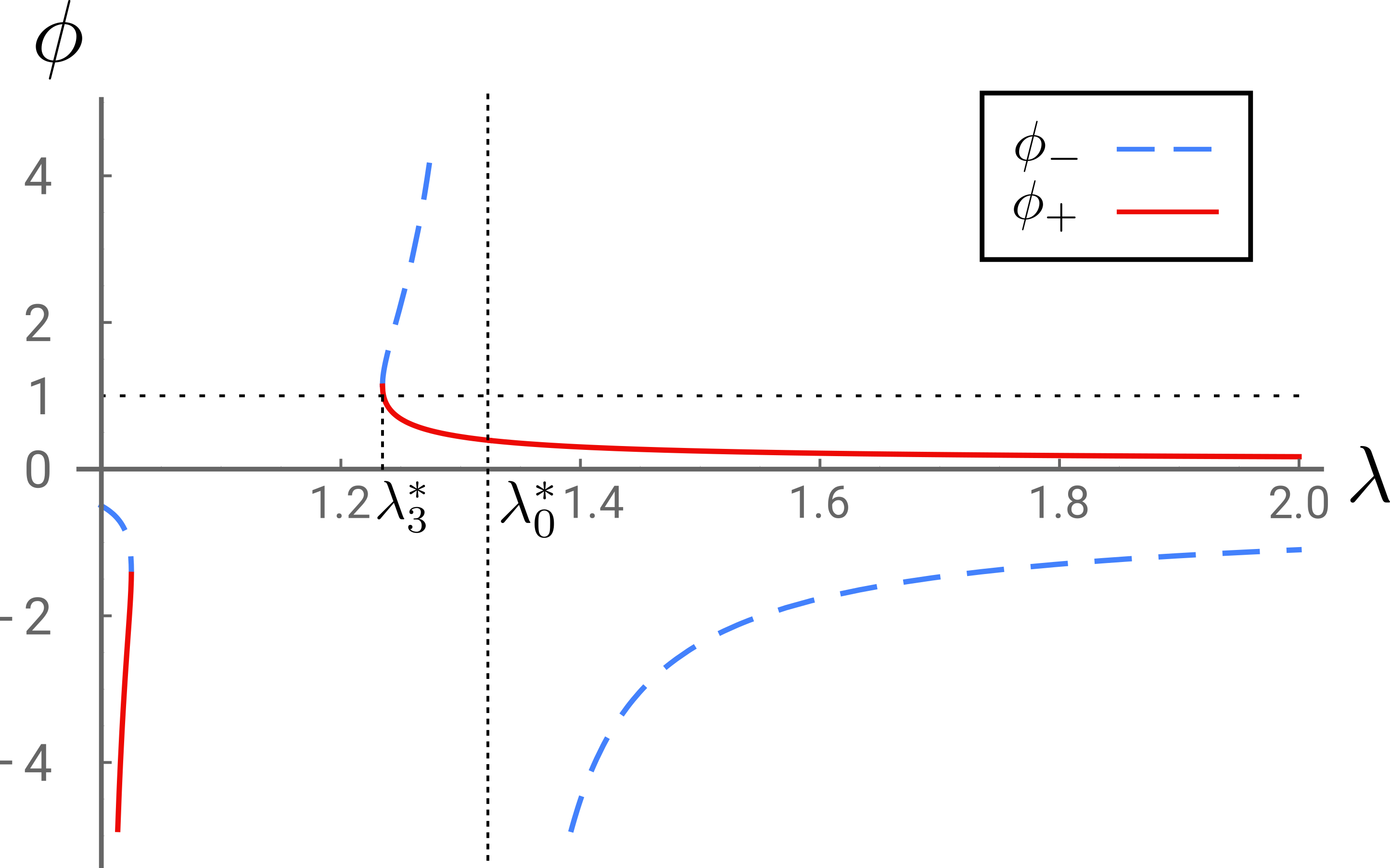}
				\end{center}
				\caption{Evolution of the solutions to the equation $p_0=1$ as a function of the outer-core diameter.
				The bounds $\lambda_1^*$ and $\lambda_2^*$ between which the solutions are not defined anymore have not been represented for clarity.}
			\label{FigPhipm}
			\end{figure}

			Another important quantity is the point $\lambda_0^*$ where $\mathcal{P}_2$ vanishes.
			Its value is given by:
			\begin{equation}
				\begin{split}
					\lambda_0^*&=\frac{1}{3}\!\left[\!\left(\!\frac{2(191+9\sqrt{451})}{5}\!\right)^ {1/3}
					\!\!\!\!\!\!\!-2^{2/3}\left(\frac{5}{191+9\sqrt{451}}\right)^{1/3} \!\!\!\!\!\!-1\right] \\
					&\simeq1.32 \,.
				\end{split}
			\end{equation}
			The evolution of $\phi_{\pm}$ is represented on Fig.~\ref{FigPhipm}.

			All in all, if
			\begin{itemize}
				\item $1\leqslant\lambda\leqslant\lambda^*_1$, the spinodal equation $p_0\leqslant1$ has a solution for any $\phi\in[\phi_+;\phi_-]$.
				However, as can be seen on the graph Fig.~\ref{FigPhipm}, both boundaries are negative, so that such values of $\phi$ can be discarded on physical grounds.

				\item $\lambda_1^*\leqslant \lambda\leqslant \lambda_2^*$, the spinodal equation has no solution.

				\item $\lambda_2^*\leqslant \lambda\leqslant \lambda_0^*$, the spinodal equation is solved for any $\phi\in[\phi_-;\phi_+]$.
				However, $\phi$ must in addition be less than 1.
				This is not realized for $\phi_+(\lambda_2^*)=\phi_-(\lambda_2^*)\simeq1.14$.
				Therefore acceptable solutions only exists if $\lambda\geqslant \lambda_3^*$, with
				\begin{equation}
					\begin{split}
						\lambda_3^*&= \frac{1}{2}\sqrt{\Upsilon^*}+\frac{1}{2}\sqrt{\frac{48}{5\sqrt{\Upsilon^*}}-\Upsilon^*} \\
						&\simeq 1.24\,,
					\end{split}
				\end{equation}
				where
				\begin{equation}
					\Upsilon^*=\frac{(38880-6480\sqrt{6})^{1/3}}{5}+\frac{2[6(6+\sqrt{6})]^{1/3}}{5^{2/3}} \,.
				\end{equation}

				\item $\lambda_0^*\leqslant \lambda$, the spinodal equation is solved for any $\phi\geqslant\phi_-(\lambda)$.
				Typical values are $\phi_-(\lambda_0^*)\simeq0.39$ and
				\begin{equation}
				\label{eqPhiInf}
					\lim_{\lambda\rightarrow+\infty}\phi_-(\lambda)=\phi_-^\infty=\frac{\sqrt{19}-3}{10}\simeq0.14\,.
				\end{equation}

			\end{itemize}

			To conclude, the low-temperature expansion of the hard-sphere solution allows for a fluid-fluid transition, but only for certain ranges of parameters.
			Firstly, this spinodal only exist for the largest possible values of $p$, that is,  moderately low temperature.
			Then, it requires a sufficiently large outer core, with typically $\delta\gtrsim 24\%$, and a packing fraction bigger than a lower bound $\phi_-$ that
			does not vanish, even for infinitely large particles.

			Once again, it should be kept in mind that the quantitative aspects of this study can be quite sensitive to the order of truncation in the $p$ expansion,
			which is all the more true that the spinodal line shows up only for moderate $p$s.
			However, the quantitative picture -- spinodal line that shows up for sufficiently high temperatures, densities and shoulder width -- should be a reliable outcome of this study.

	\subsection{High-temperature expansion}

		\subsubsection{The compressibility route to the equation of state}

			As for the low-temperature case, the compressibility equation Eq.~(\ref{eqChiEOS}) can be used to compute the equation of state.
			However, the integral in Eq.~(\ref{eqPc}) cannot be performed analytically with the high-temperature structure factor, because the expression of $\varphi$ dependence
			of $S(0)$ is too involved.
			Therefore, the compressibility equation of state associated to our high-temperature solution cannot be made explicit, which also means that we cannot build an
			explicit Carnahan-Starling-like equation, as in the low-temperature case.

\begin{figure}
				\begin{center}
					\includegraphics[scale=0.6]{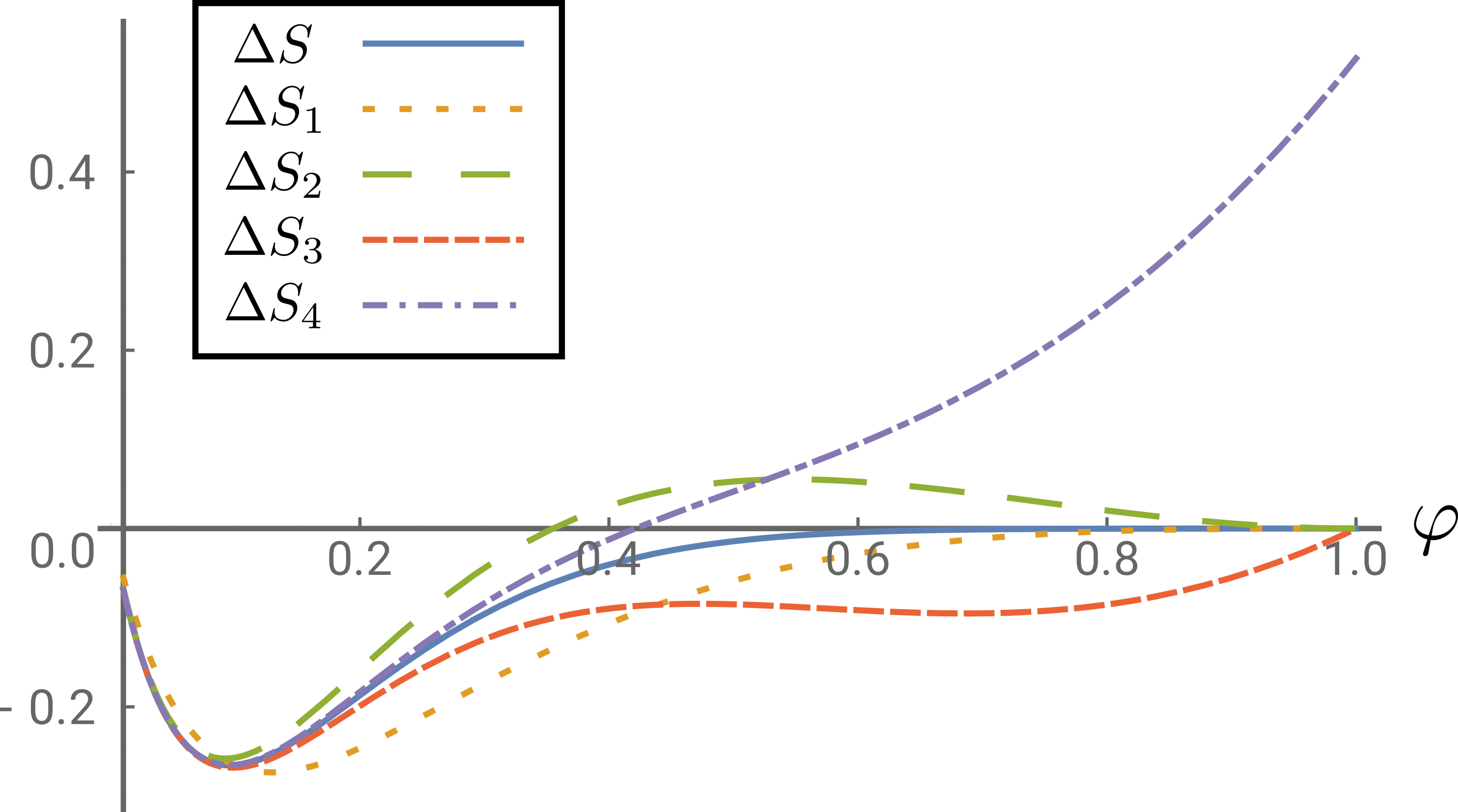}
				\end{center}
				\caption{Evolution of $\Delta S$ with the packing fraction $\varphi$, as well as some of its $\delta$-series truncations, for $\lambda=1.25$.
				The full line is the exact first order correction $\Delta S$.
				The dotted line is the first order of the $\delta$-expansion $\Delta S_1$.
				The loosely dashed line is the series truncated at order two $\Delta S_2$, the tightly dashed one is the third order truncation $\Delta S_3$.
				the dash-dotted line is the fourth order truncation $\Delta S_4$.}
			\label{FigDeltaSHT}
			\end{figure}

			One way to simplify the expressions further, and be able to perform the packing fraction integral, is to expand everything in powers of $\delta$.
			In order to do so, let us set some notations, the first correction to $S(0)$, $\Delta S$ is defined as:
			\begin{equation}
			\label{eqS0HT}
				S(0)=\frac{(1-\varphi)^4}{(1+2\varphi)^2}+\Gamma\,\Delta S\,.
			\end{equation}
			Then, the $\delta$-expansion of $\Delta S$ truncated at order $n$ is called $\Delta S_n$.
			The evolution of different $\Delta S_n$'s are represented on Fig.~\ref{FigDeltaSHT}.

			On this example, the first order truncation is quite faithful, it only underestimates a bit the exact result.
			The second order truncation, on the other hand, is getting qualitatively wrong as in a whole region the correction is positive, contrary to the exact result.
			Going to order three, only the low-packing fraction sector is getting better than first order.
			At moderate packing fractions, $\Delta S_3$ is not closer to $\Delta S$ than $\Delta S_1$.
			The order four truncation is seriously wrong except in the low-$\varphi$ regime.
			What this example illustrates is that one must be cautious when expanding in powers of $\delta$.
			Truncating the series at higher order does not automatically make the approximation better in the whole range of $\varphi$.

			It is nevertheless possible to perform the integral numerically, and build in that manner an equivalent to the expression Eq.~(\ref{eqPCSLT}).
			The so-obtained numerical high-temperature equation of state is compared to the more precise Rogers-Young data and displayed on Fig.~\ref{figPd5} and Fig.~\ref{figPd15}.
			For really small values of $\Gamma$, the agreement is quite satisfactory, although it quickly deteriorates.
			Already for $\Gamma = 0.5$ (which strictly speaking lies a bit away from the region $\Gamma\ll1$), significant deviations are present.
			Another remarkable feature is the fact that when $\Gamma$ is further increased the expansion becomes so bad that the pressure becomes negative.
			Hence, the thermodynamical quantities derived from the equation of state should be expected to be much more sensitive to higher orders in the $\Gamma$ expansion.

		\subsubsection{Low $\varphi$ behavior}

			Since the integral in Eq.~(\ref{eqPc}), runs on a finite range, it can be swapped with a small $\varphi$ expansion.
			Hence, the low-$\varphi$ sector of the high-temperature compressibility equation can be worked out without further approximation (in particular no $\delta$-expansion is required).
			In the following, we shall combine it with the results of the previously established pressure equation of state, with the same coefficients as in Eq.~(\ref{eqPCSLT}).
			In that way, we get the most precise equation of state that can be obtained from our solution.
			Its expansion reads:
			\begin{equation}
				\begin{split}
					\frac{\beta P_{HT}^{CS}}{\rho}&=1+4\big[1+(\lambda^3-1)\Gamma\big]\varphi \\
					&+\frac{\varphi^2}{18}\Big[180+\Gamma\big(15(4\lambda^2-4\lambda-1)e^{4(1-\lambda)}+8\lambda^6 \\
					&\:-252\lambda^4+480\lambda^3+144\lambda^2-365\big)\Big]+O(\varphi^3)\,,
				\end{split}
			\end{equation}
			which difference to the virial equation of state is given by
			\begin{equation}
				\frac{\beta P_{HT}^{CS}}{\rho}-\frac{\beta P^{V}}{\rho}=-\frac{32}{3}\Gamma\,\delta\,\varphi^2+O(\varphi^3,\delta^2,\Gamma^2)\,.
			\end{equation}

			All in all, despite our approximations, and the involved values of coefficients in the high-temperature expansion, a Carnahan-Starling equivalent to the equation of state can still
			be build in the low density regime.
			It matches exactly the first non-trivial virial coefficient of the square-shoulder potential, which is definitely a satisfactory test of our approximation scheme.
			The typical error is of second order in $\Gamma$, second in $\delta$ and third in $\varphi$, which is comparable to the low-temperature solution.

		\subsubsection{Thermoelastic coefficients}

			Although the compressibility equation of state does not have an explicit expression at high temperature, it can still be used to derive some of the thermodynamic properties of the
			square-shoulder fluid.
			Indeed, many of them involve derivatives, so that only the integrand in Eq.~(\ref{eqPc}) is involved.
			Hence, the effect of the appearance of a repulsive outer-core in a hard-sphere system can be discussed.
			All the following results are derived from the combination $2 P_{HT}^c/3+P_{HT}^v/3=P_{HT}^{CS}$, which is the most precise equation of state that can be built from our high-temperature solution.
			As in the low-temperature case, we will be mostly concerned by the dimensionless counterparts of thermodynamical quantities.

\begin{figure}
				\begin{center}
					\includegraphics[scale=0.5]{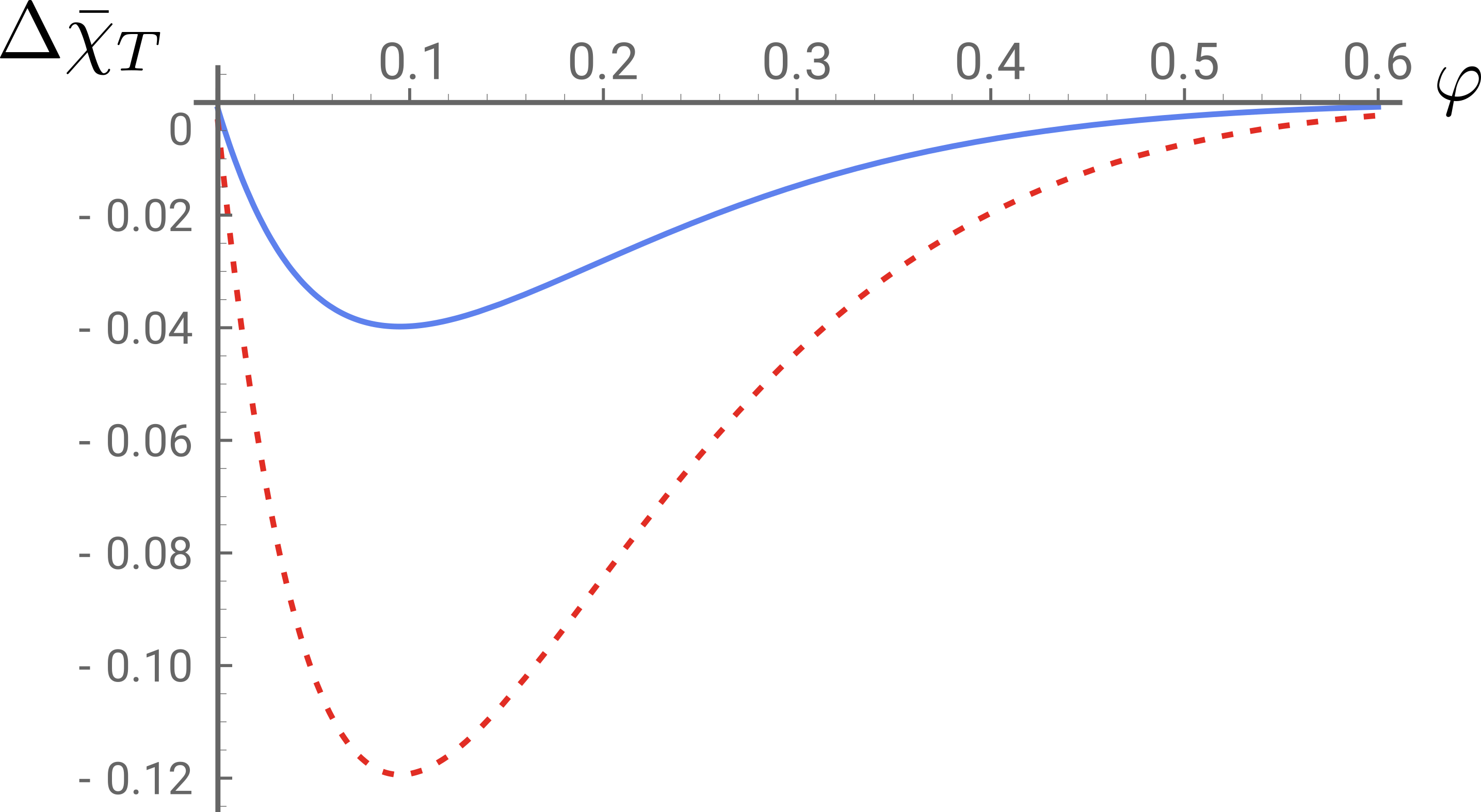}
				\end{center}
				\caption{Evolution of the dimensionless correction to the isothermal compressiblity with the inner-core packing fraction for $\lambda=1.2$.
				The full line corresponds to $\Gamma=0.2$, and the dotted one to $\Gamma=0.6$.}
			\label{FigDeltaChiHT}
			\end{figure}

			We first compute the dimensionless isothermal compressibility $\bar{\chi}_T$.
			Its analytical expression is fully explicit (no integral term remain), and has the following structure:
			\begin{equation}
				\bar{\chi}_T=\frac{(1-\varphi)^4}{\mathcal{P}_1(\varphi)}+\Delta\bar{\chi}_T\,.
			\end{equation}
			The polynomial $\mathcal{P}_1$ has been defined in Eq.~(\ref{eqPphi}); the first term in the equation is therefore the expected Carnahan-Starling result \cite{Lee95}.
			The renaming term $\Delta\bar{\chi}_T$ vanishes as $\Gamma$ goes to 0.
			Its quite lengthy expression is given in Appendix \ref{AHTTc}.
			Its evolution with $\varphi$ is also displayed on Fig.~\ref{FigDeltaChiHT} for different values of $\Gamma$.
			It appears that the isothermal compressibility is always reduced by the high-temperature correction: adding a soft outer-core to the particles reduces their compressibility.

\begin{figure}
				\begin{center}
					\includegraphics[scale=0.5]{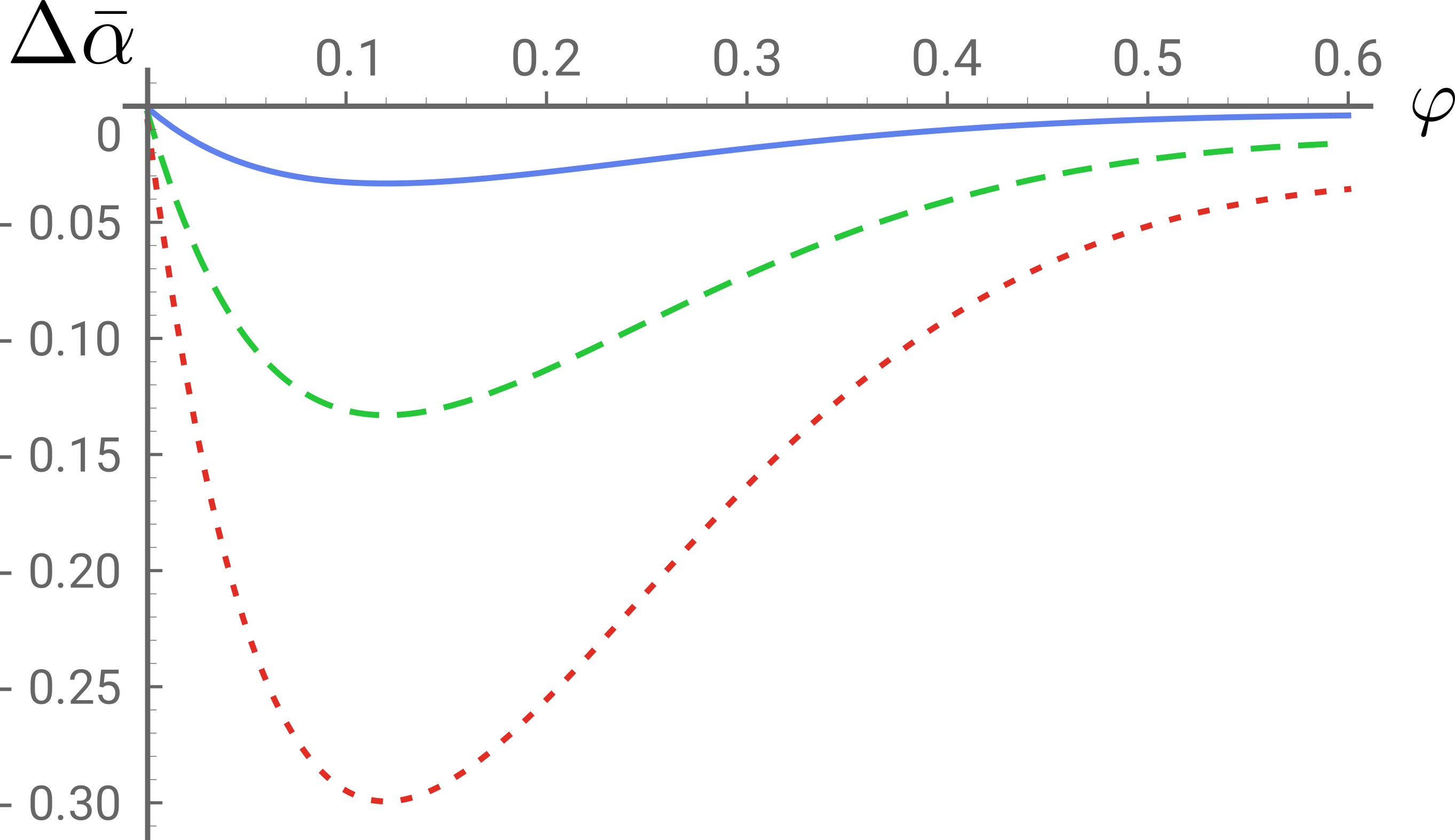}
				\end{center}
				\caption{Evolution of the dimensionless correction to the thermal expansion coefficient with the inner-core packing fraction for $\lambda=1.2$.
				The full line corresponds to $\Gamma=0.2$, the dashed one to $\Gamma=0.4$ and the dotted one to $\Gamma=0.6$.}
			\label{FigAlphaHT}
			\end{figure}

			The thermal expansion coefficient can also be computed explicitly.
			It has the expected structure:
			\begin{equation}
				\bar{\alpha}=\frac{\mathcal{P}_2(\varphi)}{T^*\mathcal{P}_1(\varphi)}+\Delta\bar\alpha\,,
			\end{equation}
			where once again, it should be understood that $\Delta\bar\alpha$ vanishes with $\Gamma$.
			The first term is the Carnahan-Starling result \cite{Wilhelm74}.
			The polynomial $\mathcal{P}_2$ is defined in Eq.~(\ref{eqPphi2}).
			The full expression of $\Delta\bar\alpha$ can be found in Appendix \ref{AHTTc}.
			Its evolution with $\varphi$ is represented on Fig.~\ref{FigAlphaHT}.
			Interestingly, this correction is negative, as in the low temperature case: from the point of view of the thermal expansion, the qualitative effect of adding a soft core or
			softening a hard core is the same.
			It is also worth noting that in the high-temperature regime, no change of sign is observed at low-densities, even when the temperature is significantly decreased.
			Indeed, the temperature expansion is now organized in powers of $\Gamma\propto 1/T$, so that no sub-leading logarithmic corrections are present in that case.
			The $\Delta\bar\alpha$ dependence on $T$ is thus much simpler than in the low-temperature regime.

			\begin{figure}
				\begin{center}
					\includegraphics[scale=0.5]{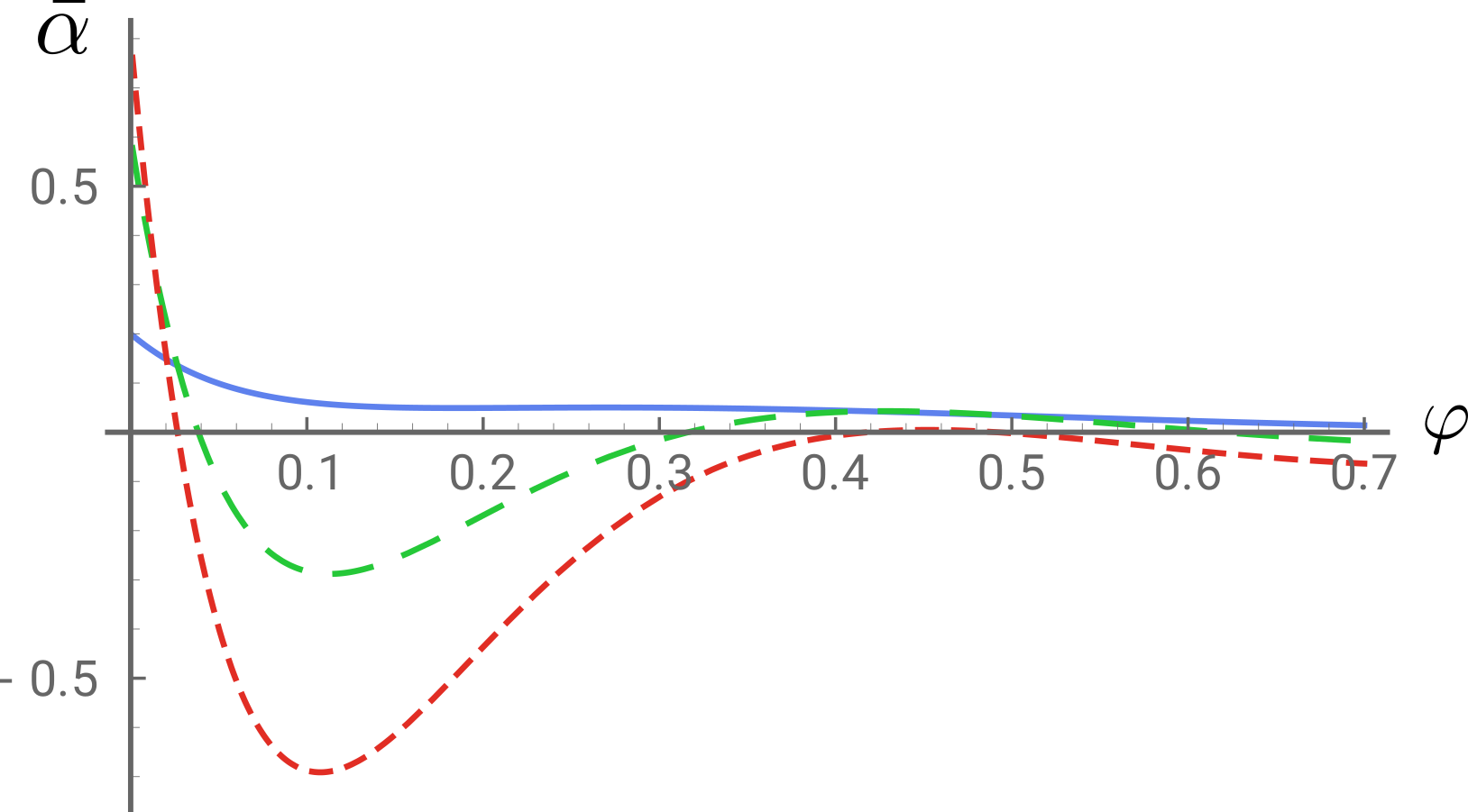}
				\end{center}
				\caption{Evolution of the dimensionless thermal expansion coefficient with the inner-core packing fraction for $\lambda=2$.
				The full line corresponds to $\Gamma=0.2$, the dashed one to $\Gamma=0.6$ and the dotted one to $\Gamma=0.8$.}
			\label{FigAlphaHT1}
			\end{figure}

\begin{figure}
				\begin{center}
					\includegraphics[scale=0.5]{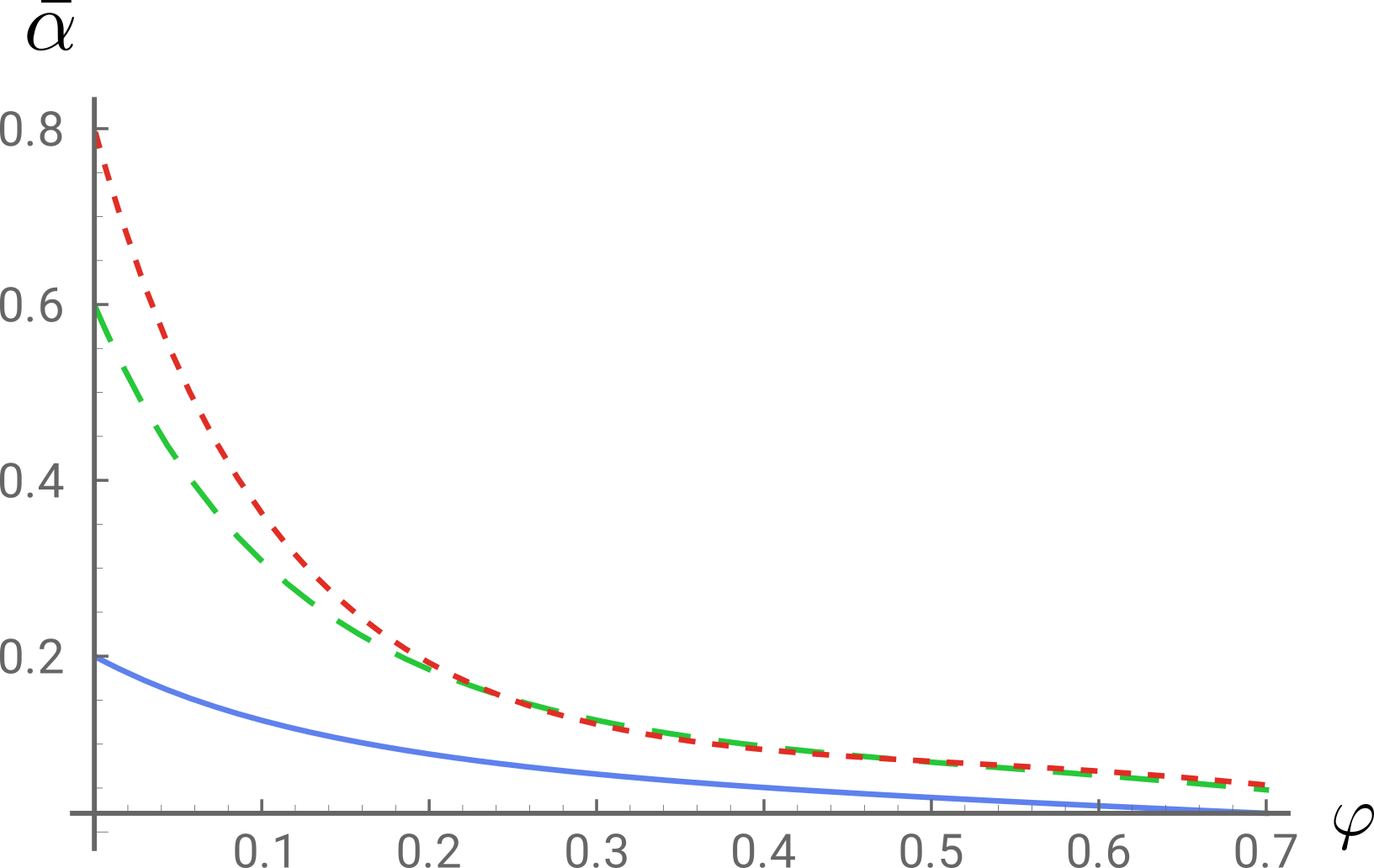}
				\end{center}
				\caption{Evolution of the dimensionless thermal expansion coefficient with the inner-core packing fraction for $\lambda=1.2$.
				The full line corresponds to $\Gamma=0.2$, the dashed one to $\Gamma=0.6$ and the dotted one to $\Gamma=0.8$.}
			\label{FigAlphaHT2}
			\end{figure}

			The full thermal expansion coefficient can show signs of abnormality (see Fig.~\ref{FigAlphaHT1} and Fig.~\ref{FigAlphaHT2}).
			As the other extreme temperature regime, it cannot stay negative up to arbitrarily high temperatures because of the hard-sphere limit,
			and it requires high enough values of $\lambda$ to be present.
			A second regime of negative $\bar\alpha$ can be observed on Fig.~\ref{FigAlphaHT}, but one must be careful since as $\varphi$ gets close to
			$0.6$, it is expected that the liquid freezes into a glass, so that our description could be a bit too sketchy in this regime.

\begin{figure}
				\begin{center}
					\includegraphics[scale=0.5]{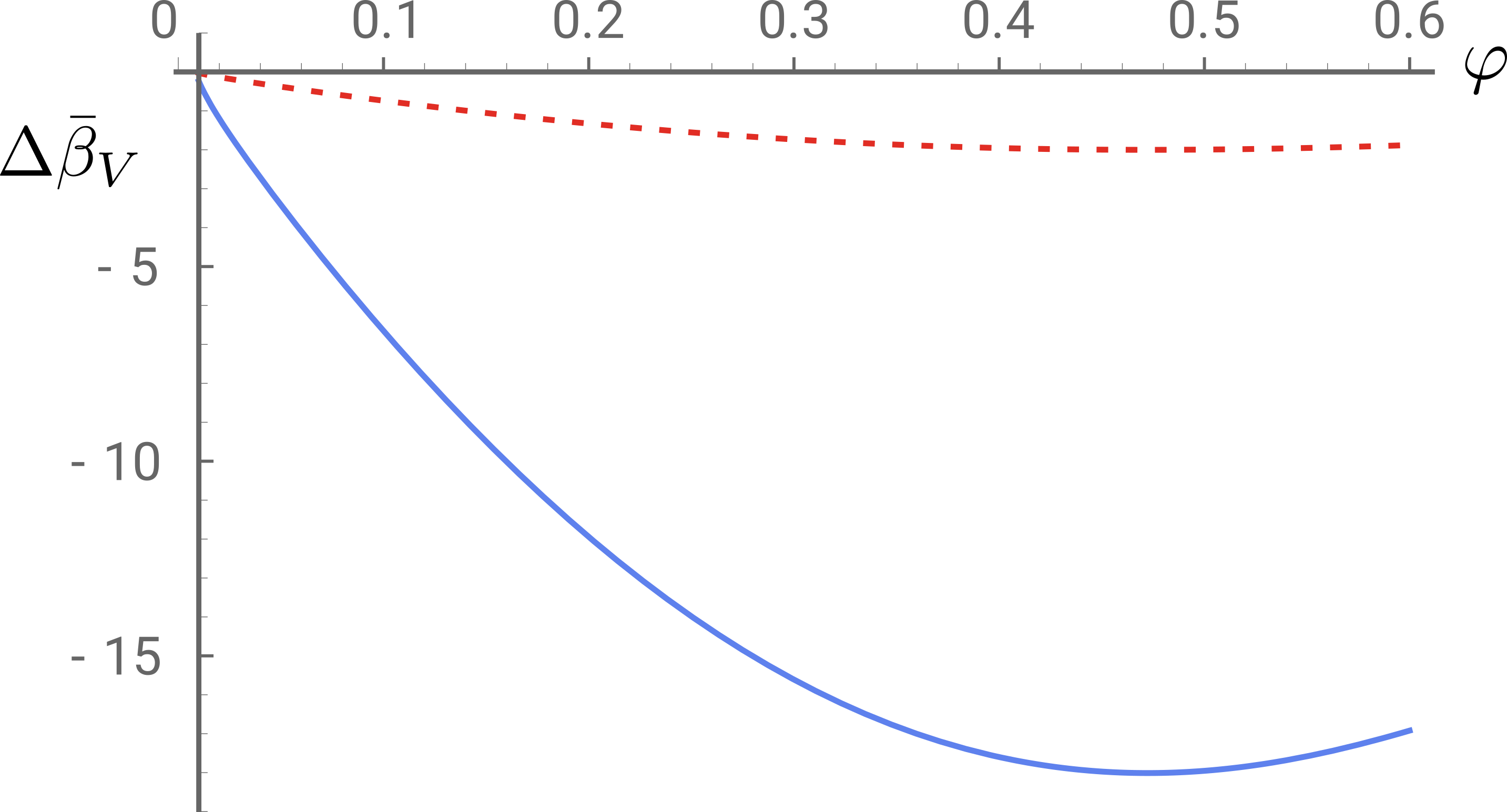}
				\end{center}
				\caption{Evolution of the dimensionless correction to the isochoric compression coefficient with the inner-core packing fraction for $\lambda=1.2$.
				The full line corresponds to $\Gamma=0.2$, and the dotted one to $\Gamma=0.6$.}
			\label{FigDeltaBetaHT}
			\end{figure}

			The isochoric compression coefficient still has the expected form:
			\begin{equation}
				\bar\beta_V=\frac{1}{T^*}+\Delta\bar\beta_V\,,
			\end{equation}
			where the full expression of $\Delta\bar\beta_V$ is given in Appendix \ref{AHTTc}.
			However, it now depends on the first correction to the integral in Eq.~(\ref{eqPc}), that is denoted $\mathcal{V}(\varphi,\lambda)$ in the Appendix \ref{AHTTc}.
			This integral can still be evaluated numerically, which yields the curves on Fig.~\ref{FigDeltaBetaHT}.
			As in the low-temperature case, the first correction to $\beta_V$ is always negative.

			Thermodynamical consistency can also be checked, the relation:
			\begin{equation}
				\alpha = P\,\beta_V\,\chi_T
			\end{equation}
			holds, independent of the form of $\mathcal{V}(\varphi,\lambda)$.

\begin{figure}
				\begin{center}
					\includegraphics[scale=0.5]{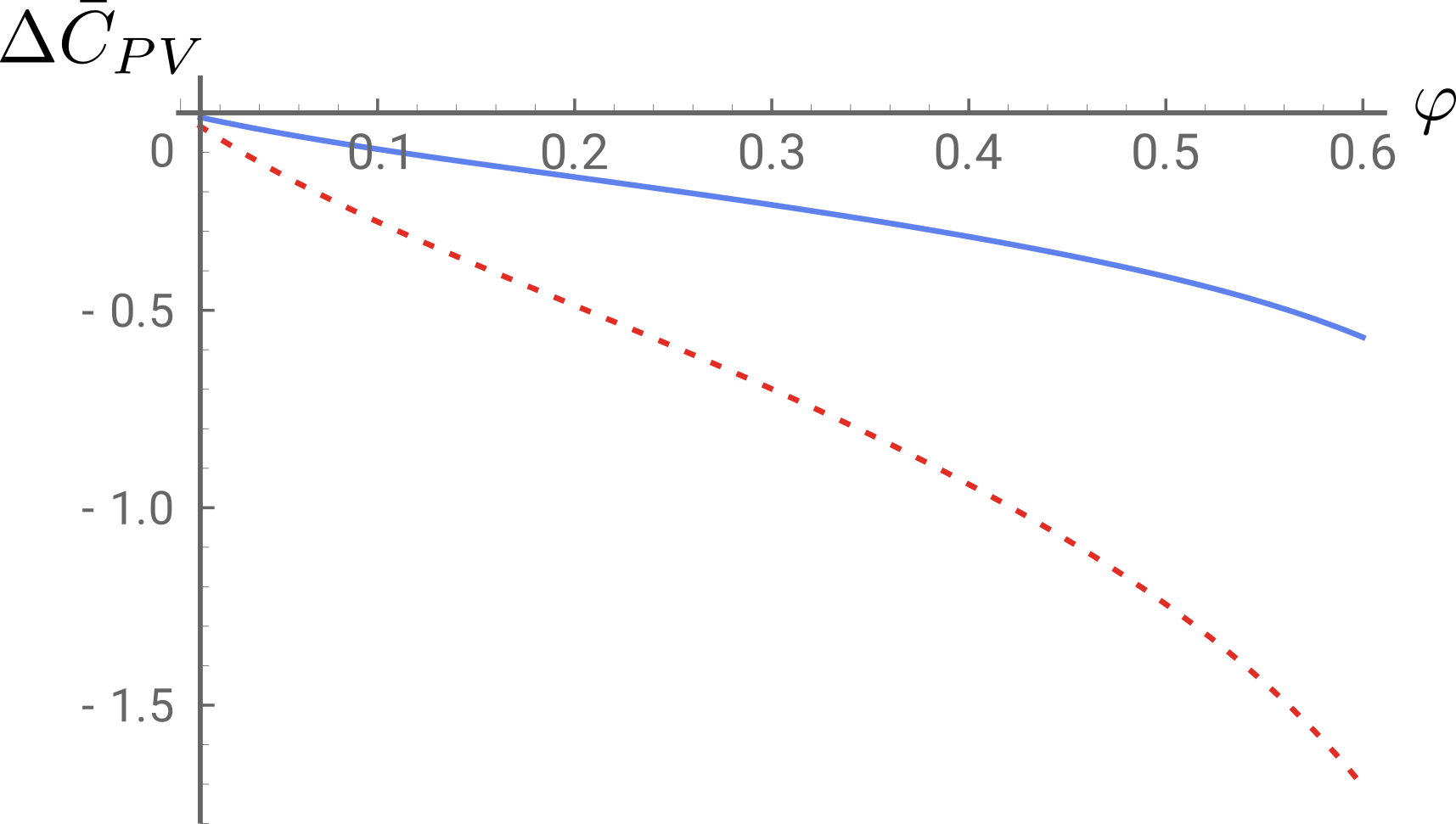}
				\end{center}
				\caption{Evolution of the dimensionless correction to the heat capacity difference with the inner-core packing fraction for $\lambda=1.2$.
				The full line corresponds to $\Gamma=0.2$, and the dotted one to $\Gamma=0.6$.}
			\label{FigDeltaCpCvHT}
		\end{figure}

			Finally, Mayer's relation can be applied to derive the heat capacity difference:
			\begin{equation}
			\label{eqcpcvHT}
				\begin{split}
					\bar{C}_P-\bar{C}_V &= (T^*)^2\frac{\bar\alpha^2}{\bar\chi_T} \\
					&=\frac{\mathcal{P}_2(\varphi)^2}{(1-\varphi)^4\mathcal{P}_1(\varphi)}+\Delta\bar{C}_{PV}\,.
				\end{split}
			\end{equation}
			This is consistent with the hard-sphere result \cite{Wilhelm74}.
			The evolution of $\Delta\bar{C}_{PV}$ is always negative, as shown on Fig.~\ref{FigDeltaCpCvHT} (but when the Carnahan-Starling contribution is added, as is Eq.~(\ref{eqcpcvHT}),
			the total result is positive).
			Its full expression is given in Appendix \ref{AHTTc}.

			All in all, even in the high-temperature regime where the compressibility equation of state is not explicit, it is remarkably possible to derive the analytical expressions of most of the
			thermodynamic quantities in a way that naturally interpolates the Carnahan-Starling's results for the hard-sphere fluid.

		\subsubsection{The fully incompressible regime}

			As in the low-temperature case, one can look for a possible spinodal line separating a low-density fluid and a high-density one.
			However, since the first correction to $1/S(0)$ is always positive, as can be seen on Fig.~\ref{FigDeltaSHT}, the equation $1/S(0)=0$ has no solution.
			Hence at this order, there is no spinodal line emerging from the high-temperature regime.
			Let us stress once again that this result is very much dependent on the order of truncation.
			Here, the spinodal equation, of type $x_0+x_1\,\Gamma=0$ could be too simplistic to allow for physical solutions to exist.

\begin{figure}
				\begin{center}
					\includegraphics[scale=0.5]{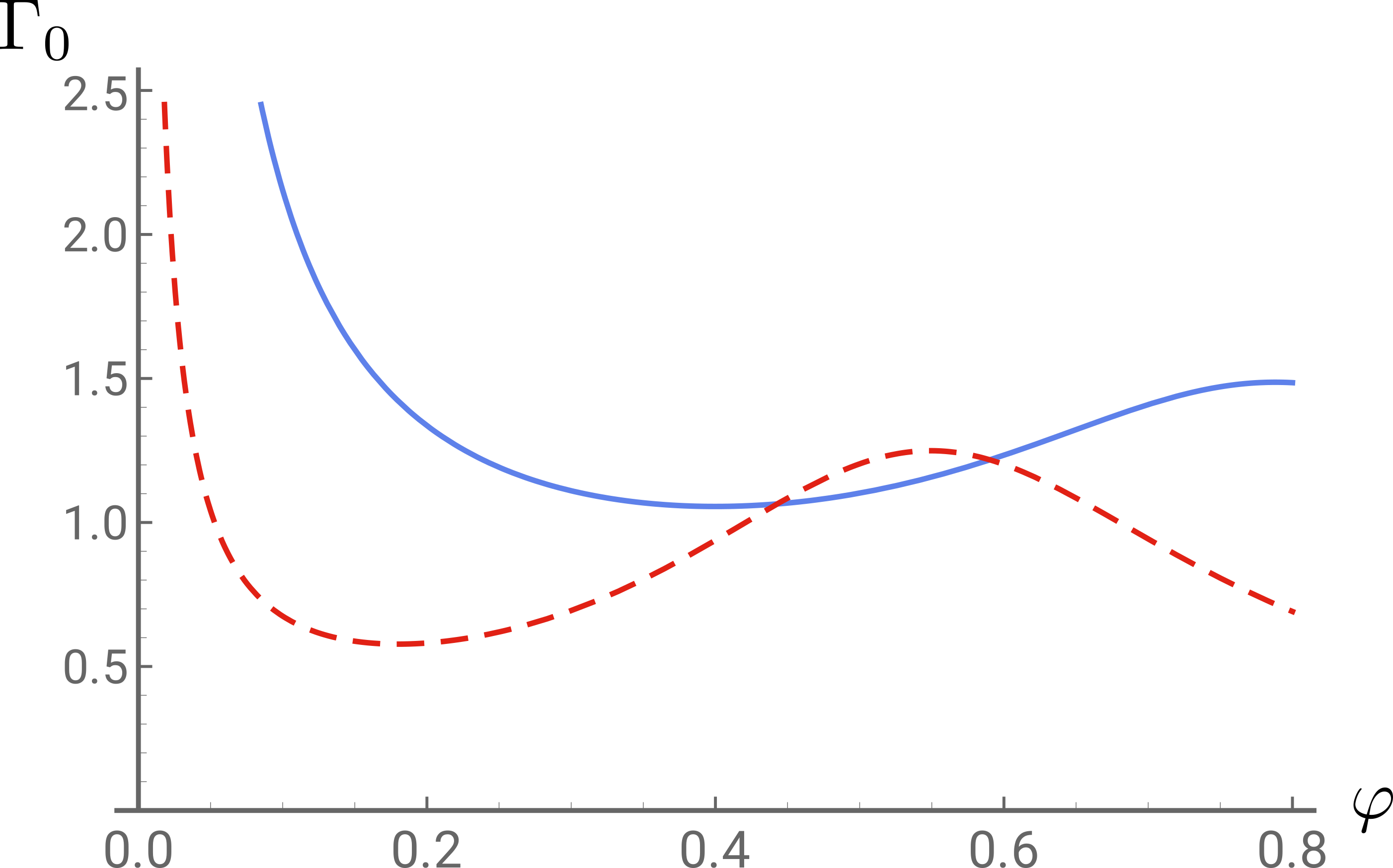}
				\end{center}
				\caption{Evolution of the solution $\Gamma_0$ to the incompressibility equation $S(0)=0$ with the inner-core packing fraction.
				The full line corresponds to $\lambda=1.2$, and the dashed one $\lambda=1.6$.}
			\label{FigGamma}
			\end{figure}

			On the other hand, and for the same reason, the equation $S(0)=0$ always has one solution, that we shall call $\Gamma_0$.
			Such solutions are displayed on Fig.~\ref{FigGamma} for different values of $\lambda$.
			They correspond to cases where the isothermal compressibility $\chi_T$ vanishes, namely, the fluid becomes completely incompressible.

			Contrary to the low-temperature case where additional conditions in $p$ had to be met the only physical constraint on $\Gamma$ is that it be positive,
			which from what we showed above is always met.
			However, as can be seen on Fig.~\ref{FigGamma}, for a fixed value of $\lambda$, $\Gamma_0$ cannot take arbitrarily small values.
			Namely, the fully incompressible regime does not extend to arbitrarily high temperatures.
			In particular, at small packing fractions $\Gamma_0$ takes higher and higher values.
			More precisely, the asymptotic behavior of $\Gamma_0$ is given by:
			\begin{equation}
				\Gamma_0\underset{\varphi\rightarrow0}{\sim}\frac{1}{8(\lambda^3-1)\varphi}\,.
			\end{equation}

			Solutions thus only involve moderate temperatures, what requires to extrapolate our solution outside of the regime in which it is
			the most precise.
			Therefore quantitative results must be taken with caution.


\section{Conclusion}

	All in all, our investigation allowed us to give a picture of the thermodynamic properties of the square-shoulder fluid.
	It gave a qualitative picture of the effect of the addition of a soft core, or softening of the hard core of a hard-sphere system.
	Further properties, such as the prospective transition between a high-density and low-density fluid, anomalies of the thermal expansion coefficient,
	or the transition to a fully incompressible regime have also been investigated.
	They reveal that subtle phenomena are at play in square-shoulder fluids at moderate temperatures and density.
	This is a strong hint that part of the exotic physics of the square-shoulder fluid is already present at the lowest order
	correction to the hard-sphere physics.
	Unfortunately our simple model is not expected to be very accurate in such regions of moderate temperatures.
	Further work is thus required to be able to describe more accurately these regions.
	If such results were to be known analytically, they could be compared to the results of the present study that must be reliable in both
	temperature limits.

\section*{Acknowledgments}

	This work was funded by a DAAD grant.
	We thank T. Kranz for careful reading of this manuscript.
	We warmly thank P. Ziherl for helpful and enlightening discussions.
	We thank N. Gadjisade for the data she provided.

\appendix
\section{Large $q$ behavior -- High temperature}
\label{AqHT}
	We recall the basic definitions:
	\begin{widetext}

	\[c_q\underset{q\rightarrow+\infty}{=}\; A(\Gamma,\varphi,\lambda)\,\frac{4\pi\,R}{q^2}+\; B_1(\Gamma,\varphi,\lambda)\,\frac{4\pi\,R}{q^2}\cos(q\,R)+\; B_2(\Gamma,\varphi,\lambda)\,\frac{4\pi\,d}{q^2}\cos(q\,d)+O(|q|^{-3})\;.\]

	\end{widetext}

	The coefficients are defined such that $A(\Gamma,\varphi,\lambda)=\Gamma\,A^{(1)}(\varphi,\lambda)+O(\Gamma^2)$, $B_1(\Gamma,\varphi,\lambda)=B_b(\varphi)+\Gamma\,B_1^{(1)}(\varphi,\lambda)+O(\Gamma^2)$,
	$B_2(\Gamma,\varphi,\lambda)=\Gamma\,B_2^{(1)}(\varphi,\lambda)+O(\Gamma^2)$.
	There are given by:

	\begin{widetext}
	\begin{equation}
		\begin{split}
			A^{(1)}(\varphi,\lambda)=& \frac{e^{-\frac{4 (6 \varphi (\varphi +1)+\lambda (\varphi^2 -2 \varphi +10))}{2 \varphi^2 -7 \varphi +5}}}
			{19440 (\varphi-1)^3 \varphi^5 (7 \varphi -10) ((\varphi -2) \varphi +10)^4 (\varphi (5 \varphi -13)-10)} \\
			&\times \bigg(3645\, e^{\frac{4 (7 \varphi^2 +4 \varphi +10)}{2 \varphi^2-7 \varphi +5}} (5-2 \varphi)^4 (\varphi (5 \varphi -13)-10) \\
			&\ \times(\varphi (\varphi (\varphi (\varphi (98 \varphi^2 -726 \varphi+2493)-4726)+5415)-3825)+2000) \varphi^8 \\
			&-90\, e^{7 (\lambda +1)-\frac{28}{\varphi-1}+\frac{5 (3\lambda +31)}{2 \varphi-5}} (5-2\varphi)^2 ((\varphi -2)\varphi +10)^4 \\
			&\ \times(\varphi (\varphi (\varphi (\varphi (\varphi (\varphi (91 \varphi -925)+2010)-1295)-604)-93)+97)-10) \varphi^3 \\
			&+90\, e^{\lambda +\frac{6 \lambda +22}{1-\varphi }+\frac{5 (3 \lambda +31)}{2 \varphi -5}+13} (5-2 \varphi)^2 (\varphi (5 \varphi -13)-10)
			\Big(\varphi \big(54 (\varphi -1) \varphi ^3 (2 \varphi +1) (7 \varphi -10) ((\varphi -2) \varphi +10)^3 \lambda^4 \\
			&\quad +36 (\varphi-1) \varphi^2 (\varphi +2) (7 \varphi -10)((\varphi -2) \varphi +10)^2 (\varphi (\varphi (5 \varphi -33)+6)-5) \lambda^3\\
			&\quad-18 \varphi (7 \varphi -10) ((\varphi -2) \varphi +10) (\varphi (\varphi (\varphi (\varphi (\varphi (\varphi (\varphi (67\varphi -290)+1747)-5075)+5899)+2500) \\
			&\quad-2471)-290)+100) \lambda^2 +6 ((\varphi -2) \varphi +10) (\varphi (\varphi (\varphi (\varphi (\varphi (\varphi (\varphi (\varphi (\varphi (1547 \varphi -5111)+48159)-218868) \\
			&\quad+477456)-299142)-141168)+55398)+19080)+4600)-1000) \lambda \\
			&\quad -\varphi (\varphi (\varphi (\varphi (\varphi (\varphi (\varphi (\varphi (\varphi (\varphi (\varphi (67249 \varphi-794470)+4465284)-14788663)+32588860)-50016681) \\
			&\quad +55377636)-37195593)+10315260)-2395066)+69820)+132600)-58000\big)+10000\Big) \varphi^3 \\
 			&+10\, e^{\frac{6 \varphi (2 \varphi +9)+2 \lambda (8 \varphi ^2-19 \varphi +20)}{2 \varphi^2 -7 \varphi +5}}(\varphi-1)^2\big((6\lambda -1) \varphi +1\big)\big((\varphi -2)\varphi +10\big)^4
			\big(\varphi(5 \varphi -13)-10\big) \\
			&\ \times(\varphi (\varphi (\varphi (\varphi (\varphi (\varphi(91 \varphi -925)+2010)-1295)-604)-93)+97)-10) \\
			& +e^{\frac{4 (6 \varphi (\varphi +1)+\lambda (\varphi^2 -2 \varphi +10))}{2 \varphi^2 -7 \varphi +5}}(\varphi -1) (7 \varphi -10) \\
			&\ \times \Big(\varphi (\varphi (3240 (\varphi -1) \varphi^4 (2 \varphi +1) ((\varphi -2) \varphi +10)^4 (\varphi (5 \varphi -13)-10) \lambda^6 \\
			&\quad -2592 (\varphi-1) \varphi^3 ((\varphi -2) \varphi +10)^4(7 \varphi^2 +\varphi+1) (\varphi (5 \varphi -13)-10) \lambda^5 \\
			&\quad -1620 \varphi^2 ((\varphi -2) \varphi +10)^3 (\varphi (5 \varphi -13)-10) (\varphi (\varphi (\varphi (\varphi (\varphi (97 \varphi -369)+480)+334)-297)-12)+10) \lambda^4 \\
			&\quad +720 \varphi ((\varphi -2) \varphi +10)^2 (\varphi (5 \varphi -13)-10)(\varphi (\varphi (\varphi (\varphi (\varphi (\varphi (\varphi (\varphi (1379\varphi-9168)+32430)-58656) \\
			&\quad +56028)+8148)-7542)-2796)-240)+100) \lambda^3 -180 ((\varphi -2) \varphi +10) (\varphi (5 \varphi -13)-10) \\
			&\quad\times(\varphi (\varphi (\varphi (\varphi (\varphi (\varphi (\varphi (\varphi  (\varphi (\varphi (13 \varphi  (841 \varphi -8223) +545112)-1676965)+3428937)-4404537)\\
			&\quad +3418440)-600201)+163836)-42688)-24780)-3600)+1000) \lambda^2 +120000 (92 \varphi +195) \\
			&\quad +\varphi^2 (\varphi (\varphi (\varphi (\varphi (\varphi (\varphi (\varphi (\varphi (2 \varphi (\varphi (\varphi (\varphi (\varphi (9098275 \varphi -151689546)+1160006526)-5312308353) \\
			&\quad +16025607747)-33070711761)+93832849245)-88781487822)+47876557851)+225720605)\\
			&\quad-23927379840)+12641616180)-3332604840)-1036421280)-237636000))+4500000)\!\!-\!\!1000000\!\Big)\!\!\bigg)\, ,
		\end{split}
	\end{equation}

	\begin{equation}
		\begin{split}
			B_1^{(1)}(\varphi,\lambda)=& \frac{ e^{-\frac{4 (\varphi (9 \varphi +8)
			+\lambda (\varphi^2 -2 \varphi +10))}{2 \varphi^2 -7 \varphi +5}}}{19440 (\varphi-1)^3 \varphi^5
			(7 \varphi-10) (\varphi^2 -2 \varphi +10)^4 (5 \varphi^2 -13 \varphi -10)}\\
			&\times \bigg(-10935 e^{\frac{8(5 \varphi^2 +3 \varphi +5)}{2 \varphi^2 -7 \varphi +5}}
			(5-2 \varphi)^6 (25 \varphi^5 -125 \varphi^4 +151 \varphi^3 +128 \varphi^2 -415 \varphi -250)
			\varphi^9 \\
			&+90\, e^{\frac{26 \varphi^2 +58 \varphi +20+2\lambda (7 \varphi^2 -17 \varphi +10)}
			{2 \varphi^2-7 \varphi +5}} (5-2\varphi)^2 (\varphi^2-2 \varphi +10)^4 \\
			&\ \times(13 \varphi^7 -5875 \varphi^6 +14376 \varphi^5-7853 \varphi^4 -2026\varphi^3
			+609 \varphi^2 +37 \varphi -10) \varphi^3\\
			&-90\,e^{\frac{38\varphi^2 +28 \varphi +20 +2\lambda (\varphi^2 -2 \varphi+10)}{2\varphi^2-7\varphi+5}}
			(5-2\varphi)^2 (5 \varphi^2 -13 \varphi -10) \big((756 \lambda^4 +2772 \lambda^3 -5418 \lambda^2
			+14286 \lambda -36499) \varphi^{13} \\
			&\ -2 (2997 \lambda^4 +14076 \lambda^3 -14706 \lambda^2 +66825 \lambda -203375)\varphi^{12} \\
			&\ +6(6777 \lambda^4 +27810 \lambda^3 -38235 \lambda^2 +148863 \lambda -341728)\varphi^{11} \\
			&\ +(-158436 \lambda^4 -684180 \lambda^3 +1007370 \lambda^2 -3589464 \lambda +5684617)\varphi^{10} \\
			&\ +2 (253854 \lambda^4 +952830 \lambda^3 -1709073 \lambda^2 +5132322 \lambda -4587338)\varphi^9 \\
			&\ -9 (118224 \lambda^4 +443376 \lambda^3 -825468 \lambda^2 +2036436 \lambda -781385)\varphi^8 \\
			&\ +6 (271620 \lambda^4 +915984 \lambda^3 -1406331 \lambda^2 +2861472 \lambda +254672)\varphi^7 \\
			&\ -3 (442800 \lambda^4 +1737360 \lambda^3 +750564 \lambda^2 -1899396 \lambda +3884953)\varphi^6 \\
			&\ -36 (4500 \lambda^4 -107100 \lambda^3 -370065 \lambda^2 +557346 \lambda -273916) \varphi^5 \\
			&\ +2 (270000 \lambda^4 -576000 \lambda^3 -3479400 \lambda^2 +744060 \lambda +793673)\varphi^4 \\
			&\ -20 (18000 \lambda^3 -19800 \lambda^2 -176580 \lambda +23471) \varphi^3
			+600 (300 \lambda^2 -120 \lambda -1001) \varphi^2 \\
			&\ -2000 (30\lambda -1) \varphi +10000\big) \varphi^3 \\
			&-10\, e^{\frac{2\varphi (12 \varphi+31)+2 \lambda (8 \varphi^2 -19 \varphi +20)}{2 \varphi^2
			-7 \varphi+5}}(\varphi-1)^2 \big((6 \lambda -1) \varphi +1\big)(\varphi^2 -2 \varphi +10)^4\\
			&\ \times(65\varphi^9 -29544 \varphi^8 +148125 \varphi^7 -167403 \varphi^6 -51801 \varphi^5
			+107913 \varphi^4 +12528 \varphi^3 -6621 \varphi^2 -240 \varphi +100) \\
			&-e^{\frac{4 (\varphi (9 \varphi +8)+\lambda (\varphi^2 -2 \varphi +10))}{2 \varphi^2-7 \varphi +5}}
			(7 \varphi^2 -17 \varphi +10) \\
			&\times\big(10 (3240 \lambda^6 -1296 \lambda^5 -88290 \lambda^4 +381960 \lambda^3 -652950 \lambda^2
			+727205) \varphi^{18} \\
			&\ -12(29970 \lambda^6 -4968 \lambda^5 -910170 \lambda^4 +4152780 \lambda^3 -7745250 \lambda^2
			+9247363) \varphi^{17} \\
			&\ +12 (236520 \lambda^6-14256 \lambda^5 -6905925 \lambda^4 +30444120 \lambda^3 -55886715 \lambda^2
			+61704296) \varphi^{16} \\
			&\ -30 (495180 \lambda^6 +59184 \lambda^5 -14231160 \lambda^4 +59669328 \lambda^3 -103502454 \lambda^2
			+91484027) \varphi^{15} \\
			&\ +12 (4981500 \lambda^6 +1453032 \lambda^5 -134414775 \lambda^4 +520128720 \lambda^3
			-821818950 \lambda^2 +475049483) \varphi^{14} \\
			&\ -12 (15260400 \lambda^6 +8075376 \lambda^5 -380042955 \lambda^4 +1302018840 \lambda^3
			-1778104530 \lambda^2 +385117384) \varphi^{13} \\
			&\ +9 (46638720 \lambda^6 +41043456 \lambda^5 -1034475840 \lambda^4 +2934959840 \lambda^3
			-3144996700 \lambda^2 -767414941)\varphi^{12} \\
			&\ -12 (60283440 \lambda^6 +83115072 \lambda^5 -1075705920 \lambda^4 +2029517640 \lambda^3
			-1068168315 \lambda^2 -1846059662) \varphi^{11} \\
			&\ +3 (237168000 \lambda^6 +690571008 \lambda^5 -2712547440 \lambda^4 -2264837760 \lambda^3
			+8800066260 \lambda^2 -6471065803) \varphi^{10} \\
			&\ -5 (60912000 \lambda^6 +523376640 \lambda^5 +1102206528 \lambda^4 -8805479616 \lambda^3
			+9189146232 \lambda^2 +1353711563) \varphi^9 \\
			&\ -120 (6480000 \lambda^6 -18230400 \lambda^5 -141056100 \lambda^4 +338124168 \lambda^3
			-127383837 \lambda^2 -231030331) \varphi^8 \\
			&\ +60 (8100000 \lambda^6 +16848000 \lambda^5 -24408000 \lambda^4 -422971200 \lambda^3
			+545066784 \lambda^2 -376608625) \varphi^7 \\
			&\ +120 (2700000 \lambda^6 -14040000 \lambda^5 -38340000 \lambda^4 +112584000 \lambda^3
			-8082600 \lambda^2 +31390829) \varphi^6 \\
			&\ -480 (540000 \lambda^5 -1856250 \lambda^4 -7785000 \lambda^3 +9140250 \lambda^2+1949411) \varphi^5 \\
			&\ +12000 (13500 \lambda^4 -27000 \lambda^3 -84600 \lambda^2 -58433) \varphi^4
			-360000 (200 \lambda^3 -175 \lambda^2 -354) \varphi^3 \\
			&\ +1200000 (15 \lambda^2 +52) \varphi^2-1500000 \varphi -1000000\big)\bigg)\, ,
		\end{split}
	\end{equation}

	\begin{equation}
		\begin{split}
			B_2^{(1)}(\varphi,\lambda)=&-1+ \frac{\varphi (2 \varphi -5) \exp \left(-\frac{2 (\lambda -1)
			(\varphi^2 -2 \varphi +10)}{2 \varphi^2 -7 \varphi +5}\right)}{2 \lambda (\varphi -1)^2}\,.
		\end{split}
	\end{equation}
	\end{widetext}

\section{High Temperature -- Thermoelastic coefficients}
\label{AHTTc}

	\begin{itemize}

		\item Isothermal compressibility $\bar{\chi}_T=\bar{\chi}_T^{CS}+\Delta\bar{\chi}_T$

		\begin{widetext}
		\begin{equation}
			\begin{split}
				\Delta\bar\chi_T&= \frac{\Gamma (\varphi-1)^3 }{4860 (10-7 \varphi)^2 (5-2 \varphi) \varphi^4 (-5 \varphi^2+13 \varphi +10)^2 (\varphi^2 -2 \varphi +10)^5 (\varphi ^4-4 \varphi^3 +4 \varphi^2 +4 \varphi +1)^2} \\
				&\times\Big[10935\,e^{-\frac{4 (\lambda -1) (\varphi^2 -2 \varphi +10)}{(\varphi -1) (2 \varphi -5)}} (2\varphi -5)^5 (-5 \varphi^2 +13 \varphi +10)^2 
				\big(140 \varphi^{10} +4 (35 \lambda -426) \varphi^9 +(584 \lambda +12373) \varphi^8 \\
				&\: -7 (1108 \lambda +9257) \varphi^7 +(41876 \lambda +205945) \varphi^6 -(123280 \lambda +380309) \varphi^5 +5 (38296 \lambda+72227) \varphi^4 \\
				&\:-25 (2752 \lambda +3971) \varphi^3-1000 (208 \lambda +101) \varphi^2 +1250 (160 \lambda +133) \varphi-125000\big) \varphi^9 \\
				&-90\,e^{\frac{2 (\lambda -1) (5 \varphi^2 -13 \varphi -10)}{(\varphi -1) (2 \varphi -5)}} (2 \varphi -5)(\varphi ^2-2 \varphi +10)^5
				\big(1820 \varphi^{13} -566906 \varphi^{12} +6099741 \varphi^{11} -19583639 \varphi^{10} \\
				&\:+2071649 \varphi^9 +97809633 \varphi^8 -166298010 \varphi^7 +63365784 \varphi^6 +27473013 \varphi^5 +24278599 \varphi^4 -36623635 \varphi^3 \\
				&\:-688725 \varphi^2 +1316000 \varphi-6 \lambda (455 \varphi^{12}-212008 \varphi^{11}+3413720 \varphi^{10} -19088091 \varphi^9 +47826858 \varphi^8 -51636894 \varphi^7 \\
				&\:+5945901 \varphi^6 +28843728 \varphi^5-13090710 \varphi^4-3216605\varphi^3 +935950 \varphi^2 +56500 \varphi -15000)-72500\big) \varphi^4 \\
				&+90\,e^{-\frac{2(\lambda -1) (\varphi^2 -2 \varphi +10)}{(\varphi-1) (2 \varphi -5)}} (-5 \varphi^2 +13 \varphi +10)^2 \mathcal{F}_0^\chi(\varphi,\lambda)\\
				&+10 e^{\frac{6 (\lambda -1) \varphi}{\varphi -1}} (2 \varphi -5) (\varphi^2 -2\varphi +10)^5 (5 \varphi^3 -18 \varphi^2 +3 \varphi +10)^2 \big(91 (6 \lambda -1) \varphi ^{11}-3(364 \lambda^2 +55180 \lambda -9227)\varphi^{10} \\
				&\:+5 (99012 \lambda ^2+78066 \lambda -18529) \varphi^9 +(-1912584 \lambda^2+64476 \lambda +54309) \varphi^8 +6 (397462 \lambda ^2-82584 \lambda +15555) \varphi^7 \\
				&\:-6 (128696 \lambda^2 +31794 \lambda +8465) \varphi^6 -3 (98092 \lambda^2 -174810 \lambda +39733) \varphi^5 +3 (23324 \lambda^2- 35412 \lambda +35037) \varphi^4 \\
				&\:+3(1760 \lambda ^2+222 \lambda -5939) \varphi^3+(-1200 \lambda ^2+4440\lambda -629) \varphi^2-120 (5 \lambda -7) \varphi -100\big) \\
				&+(10-7 \varphi)^2 (\varphi -1) (2 \varphi^2 -7 \varphi+5)\mathcal{F}_1^\chi(\varphi,\lambda)  \\
				&-\frac{2 e^{-\frac{-8 \lambda \varphi^2+50\varphi^2 +22 \lambda \varphi +2 \varphi +40 \lambda +20}{2 \varphi^2 -7 \varphi+5}} (2 \varphi-5) (2 \varphi +1)
				(\varphi^3-3 \varphi^2 +12 \varphi -10)^5(35 \varphi^3 -141 \varphi^2 +60 \varphi +100)^2}{(\varphi-1)^5
				(7 \varphi -10)(\varphi^2 -2 \varphi +10)^4 (5 \varphi^2 -13 \varphi -10)} \\
				&\times \big(-3645 e^{-\frac{6 ((2\lambda -9) \varphi ^2-5 \lambda \varphi +\varphi -10)}{(\varphi -1) (2 \varphi -5)}} (5-2 \varphi )^6(40 \varphi ^5-224 \varphi ^4+397 \varphi ^3-139 \varphi ^2-460 \varphi -100) \varphi^8 \\
				&+90\,e^{\frac{40 \varphi^2+28 \varphi +2 \lambda(\varphi ^2-2 \varphi +10)+40}{2 \varphi ^2-7 \varphi +5}} (5-2 \varphi )^2
				(\varphi ^2-2 \varphi +10)^4(13 \varphi^6-1404 \varphi^5+3693 \varphi^4-2576 \varphi^3-153 \varphi^2 \\
				&\:+204 \varphi -20) \varphi^3 -90\,e^{\frac{52\varphi^2-2 \varphi +\lambda (-10 \varphi ^2+26 \varphi +20)+40}{2 \varphi ^2-7 \varphi +5}} (5-2 \varphi )^2(5\varphi^2-13 \varphi -10)\mathcal{F}_2^\chi(\varphi,\lambda) \\
				&-10\,e^{\frac{38 \varphi^2+32 \varphi +4 \lambda(\varphi ^2-2 \varphi +10)+20}{2 \varphi ^2-7 \varphi +5}} (\varphi -1)^2 ((6 \lambda -1) \varphi +1)(\varphi ^2-2\varphi +10)^4
				(65 \varphi ^8-7189 \varphi ^7+36587 \varphi ^6-46849 \varphi ^5 \\
				&\:-4207 \varphi ^4+28769 \varphi ^3-1222\varphi ^2-1780 \varphi +200)
				-e^{\frac{-8 \lambda \varphi^2 +50 \varphi^2 +22 \lambda \varphi +2 \varphi +40 \lambda +20}{2 \varphi^2-7 \varphi +5}} (7 \varphi ^2-17 \varphi +10) \mathcal{F}_3^\chi(\varphi,\lambda)\big)\Big]
			\end{split}
		\end{equation}

		\begin{equation}
			\begin{split}
				\mathcal{F}_0^\chi(\varphi,\lambda)&=\big(-2043944 \varphi^{20} +47693776 \varphi^{19} -576229700 \varphi^{18} +4742918049 \varphi^{17} \\
				&-28699098981 \varphi ^{16}+130215493290 \varphi ^{15}-445352281374 \varphi ^{14}+1149865256949 \varphi ^{13}-2234003281311 \varphi ^{12} \\
				&+3218122561237 \varphi ^{11}-3283506838853 \varphi ^{10}+2019480513043 \varphi ^9-75562914099 \varphi ^8-1268840559270 \varphi ^7 \\
				&+1374807798600 \varphi ^6-723936336000 \varphi ^5+108 \lambda^5 (10-7 \varphi )^2 (\varphi ^2-2 \varphi +10)^4 (4 \varphi^5 +28 \varphi^4 -197 \varphi^3 +275 \varphi^2 \\
				&-10 \varphi -100) \varphi^4 +125992762500 \varphi ^4+108 \lambda ^4 (10-7 \varphi )^2 (\varphi^2 -2 \varphi +10)^3 (8 \varphi^9 -84 \varphi^8 +628 \varphi^7 \\
				& -3454 \varphi^6 +10188 \varphi^5 -15671 \varphi^4 +14690 \varphi^3-9780 \varphi^2 +1975 \varphi +1500) \varphi ^3+51521700000 \varphi ^3 \\
				& +36 \lambda ^3 (-7 \varphi ^3+24 \varphi ^2-90 \varphi +100)^2 (88 \varphi ^{12}-1478 \varphi ^{11}+10597 \varphi ^{10}-55875 \varphi ^9+195810 \varphi ^8-393888 \varphi^7 \\
				&+443334 \varphi ^6 -485952 \varphi^5+1034355 \varphi^4 -1458790 \varphi^3+877250 \varphi^2-202000 \varphi-22500) \varphi^2-5725250000 \varphi ^2 \\
				&-18 \lambda ^2 (16856 \varphi ^{19}-350532 \varphi ^{18}+3657898 \varphi ^{17}-26145499 \varphi ^{16}+138401127 \varphi ^{15}-588197765 \varphi ^{14} \\
				&2093287835 \varphi ^{13}-6297079038 \varphi ^{12}+15854111980 \varphi^{11}-32210691365 \varphi ^{10}+49883762895 \varphi ^9-55451189800 \varphi ^8 \\
				&+42600227440 \varphi ^7-18649288800 \varphi^6-15957322000 \varphi ^5 +49846300000 \varphi ^4-46869375000 \varphi ^3 \\
				&+17183000000 \varphi ^2-1747500000 \varphi -150000000)\varphi -5855000000 \varphi \\
				&+6 \lambda(133336 \varphi ^{20}-3122606 \varphi ^{19}+36311211 \varphi ^{18}-281971353 \varphi ^{17}+1640770449 \varphi ^{16}-7628913087 \varphi ^{15} \\
				& +29116269846 \varphi ^{14}-91690824645 \varphi ^{13}+236066868444 \varphi^{12}-485382657977 \varphi ^{11}+767547842380 \varphi ^{10} \\
				&-882756775338 \varphi ^9+671061848790 \varphi ^8-246177964680 \varphi^7-97818155400 \varphi ^6+187174764000 \varphi ^5 \\
				&-84207825000 \varphi ^4-7594500000 \varphi ^3+9911500000 \varphi ^2-230000000 \varphi -75000000)+362500000\big) \varphi^4 
			\end{split}
		\end{equation}

		\begin{equation}
			\begin{split}
				\mathcal{F}_1^\chi(\varphi,\lambda)&=50 (3240 \lambda ^6-1296 \lambda ^5-88290 \lambda ^4+381960 \lambda ^3-652950 \lambda ^2+727205) \varphi^{23} \\
				&-30 (92880 \lambda ^6-27792 \lambda ^5-2655630 \lambda ^4+11764920 \lambda ^3-20991850 \lambda ^2+24207499) \varphi^{22} \\
				&+4 (7114230 \lambda ^6-1608552 \lambda ^5-207291555 \lambda ^4+923758020 \lambda ^3-1677771825 \lambda ^2+1939817327)\varphi^{21} \\
				&-4 (51103710 \lambda ^6-7604712 \lambda ^5-1508079465 \lambda ^4+6729616620 \lambda ^3-12389160315 \lambda^2+14238094867) \varphi^{20} \\
				&+30 (37365732 \lambda ^6-2449728 \lambda ^5-1102950594 \lambda ^4+4868521488 \lambda^3-8959466310 \lambda ^2+9891532889) \varphi^{19} \\
				&-2 (2444241420 \lambda ^6+67518144 \lambda ^5-71249766750 \lambda^4+305915026320 \lambda^3-548782845570 \lambda ^2 \\
				&+538192686241) \varphi^{18} +3 (5721861600 \lambda ^6\!+804902976 \lambda^5\!\!-\!162367144380 \lambda ^4\!+664824437280 \lambda ^3\! \\
				&-\!1129375577940 \lambda ^2 +866217733099) \varphi^{17} -9 (5429212560\lambda ^6+1525790976 \lambda ^5-146841124500 \lambda ^4 \\
				&+556559206080 \lambda ^3  -861579974780 \lambda ^2+407424320319) \varphi^{16} +3 (37046972160 \lambda ^6+17546905728 \lambda ^5 \\
				&-929589283440 \lambda ^4 +3112263937440 \lambda ^3-4126903273920 \lambda^2+398902975307) \varphi^{15} -5 (39180703104 \lambda ^6 \\
				&+29955087360 \lambda ^5 -867173228568 \lambda ^4+2318644054368\lambda ^3-2272495674960 \lambda ^2-1164507840887) \varphi^{14} \\
				&+3 (82695461760 \lambda^6 +107286907392 \lambda^5-1437021014400 \lambda ^4+1967894966400 \lambda ^3+94464073120 \lambda ^2 \\
				&-3670077618787) \varphi^{13}+(-170684236800\lambda ^6-508306945536 \lambda ^5+1031322235680 \lambda ^4+8542454840640 \lambda ^3 \\
				&-15561444736680 \lambda ^2+6242014295921)\varphi^{12} -20 (2565626400 \lambda ^6-25502508288 \lambda ^5-196263212544 \lambda ^4 \\
				&+937693558944 \lambda^3 -839648848344\lambda ^2-226006522423) \varphi^{11} +30 (8921880000 \lambda ^6-5725025280 \lambda ^5 \\
				&-171146921760 \lambda^4 +330178944960 \lambda^3 -17027914368 \lambda ^2-274192649227) \varphi^{10} -40 (4487400000 \lambda^6 \\
				&+9603792000 \lambda^5 -13020247800\lambda ^4-207872179200 \lambda ^3+287567949528 \lambda ^2-58443955459) \varphi^9 \\
				&-120 (920700000 \lambda^6 -4279680000\lambda ^5-23531445000 \lambda ^4+84162420000 \lambda ^3-41505208260 \lambda ^2-25680982019) \varphi^8 \\
				& +1440 (58500000\lambda ^6+19320000 \lambda ^5-327150000 \lambda ^4-1452000000 \lambda ^3+3009336550 \lambda ^2-1882874127) \varphi^7 \\
				&+2400 (13500000 \lambda ^6-76320000 \lambda ^5-165172500 \lambda ^4+720360000 \lambda ^3-181670250 \lambda ^2+105597017) \varphi^6 \\
				&-4000 (4320000 \lambda ^5-17010000 \lambda ^4-23040000 \lambda ^3+85275000 \lambda ^2+23368643) \varphi^5+300000(18000 \lambda ^4 \\
				&-48000 \lambda ^3-90380 \lambda ^2-14683) \varphi^4+8000000 (105 \lambda ^2+346) \varphi^3-1000000 (600 \lambda ^2+2461) \varphi^2 \\
				&+60000000 \varphi +100000000
			\end{split}
		\end{equation}

		\begin{equation}
			\begin{split}
				\mathcal{F}_2^\chi(\varphi,\lambda)&=\big((378 \lambda^4 +1260 \lambda^3 -3150 \lambda^2 +6726 \lambda -20245) \varphi^{12}
				-3(1062 \lambda^4 +4464 \lambda^3 -7170 \lambda^2 +20912 \lambda -80567) \varphi^{11} \\
				&+6 (3654 \lambda^4 +13428 \lambda^3-26919 \lambda^2 +72236 \lambda -222189) \varphi^{10} -2 (45090 \lambda^4 +168318 \lambda^3 -376596 \lambda^2 \\
				&+928980\lambda -2125733) \varphi^9+9 (33216 \lambda^4 +106008 \lambda^3 -295206 \lambda^2 +643148 \lambda -964925) \varphi^8 -18 (37860 \lambda^4 \\
				& +111984 \lambda^3 -360063 \lambda^2 +671954 \lambda -641292) \varphi^7 +3 (385200 \lambda^4 +938880 \lambda^3 -3373668 \lambda^2 +5337048 \lambda \\
				&-3319589) \varphi^6 -36 (34500 \lambda^4 +72300 \lambda^3 -206655 \lambda^2 +226202 \lambda -102809) \varphi^5 +72 (7500 \lambda^4 +25500 \lambda^3 \\
				& -2925 \lambda^2 -44980 \lambda +10191) \varphi^4-160 (4500 \lambda^3 +9675 \lambda^2-9870 \lambda -4231) \varphi^3
				+1200 (300 \lambda^2 +530 \lambda -281) \varphi^2 \\
				&-6000 (20 \lambda +21) \varphi +20000\big)\varphi^3 
			\end{split}
		\end{equation}

		\begin{equation}
			\begin{split}
				\mathcal{F}_3^\chi(\varphi,\lambda) &=10 (1620 \lambda^6 -1296 \lambda^5 -44550 \lambda^4 +200520 \lambda^3 -346770 \lambda^2 +443705) \varphi^{17} -2 (93960 \lambda^6 -62208 \lambda^5 \\
				&-2871450 \lambda^4 +13671360 \lambda^3 -25718310 \lambda^2+36425363) \varphi^{16} +4 (378270 \lambda^6 -227448 \lambda^5 -11209590 \lambda^4 \\
				&+52126380 \lambda^3 -97114815 \lambda^2 +134693303) \varphi^{15} -4 (2046060 \lambda^6 -1031616 \lambda^5 -59803110 \lambda^4 +267733800 \lambda^3 \\
				&-475849665 \lambda^2 +584842171) \varphi^{14}+4 (8495280 \lambda^6 -3522528 \lambda^5 -235477530 \lambda^4 +992313540 \lambda^3 -1633492575 \lambda^2 \\
				&+1621901438) \varphi^{13} -4 (27138240 \lambda^6 -8118144\lambda^5 -702588330 \lambda^4 +2711876940 \lambda^3 -3992650245 \lambda^2 \\
				& +2909867434) \varphi^{12}+10 (26415072 \lambda^6 -3763584 \lambda^5 -623530224 \lambda^4 +2134579824 \lambda^3 -2691937350 \lambda^2 \\
				&+1285613177) \varphi^{11} -4 (123444000 \lambda^6 +6905088 \lambda^5 -2495838420 \lambda^4 +7044079680 \lambda^3 -7024662225 \lambda^2 \\
				&+1844967967) \varphi^{10} +5 (120528000 \lambda^6 +61585920 \lambda^5 -1961348256 \lambda^4 +3642478848 \lambda^3 -2081782296 \lambda^2 \\
				&+247814725) \varphi^9 -10 (45360000 \lambda^6 +60134400 \lambda^5 -314960400 \lambda^4 -618929568 \lambda^3 +1194566364 \lambda^2 \\
				&+184924679) \varphi^8 -80 (2025000 \lambda^6 -10692000 \lambda^5 -75168000 \lambda^4 +268399800 \lambda^3 -185576751 \lambda^2 -66979673) \varphi^7 \\
				&+400 (810000 \lambda^6 -8626500 \lambda^4 +1634400 \lambda^3 +11216340 \lambda^2 -11624801) \varphi^6 -320 (1620000 \lambda^5 +1012500 \lambda^4 \\
				&-12442500 \lambda^3 +2968875 \lambda^2 +531958) \varphi^5+8000 (40500 \lambda^4 +36000 \lambda^3 -155925 \lambda^2 -55939) \varphi^4 \\
				&-20000 (7200 \lambda^3 +5400 \lambda^2 +3991) \varphi^3 +100000 (360 \lambda^2 +553) \varphi^2 +10000000 \varphi -2000000
			\end{split}
		\end{equation}

		\end{widetext}

		\item Thermal expansion coefficient $\bar{\alpha}=\bar\alpha^{CS}+\Delta\bar\alpha$

		\begin{widetext}
		\begin{equation}
			\begin{split}
				\Delta\bar\alpha&=-\frac{e^{-\frac{2(2 \varphi (3 \varphi +2)+\lambda (6 \varphi^2 +15 \varphi +20))}{2 \varphi^2 -7 \varphi +5}} (\varphi^3 -\varphi^2 -\varphi -1)}
				{4860 (T^*)^2 (10-7 \varphi )^2 \varphi^4 (2 \varphi -5) (-5 \varphi^2 +13 \varphi +10)^2 (\varphi^2 -2 \varphi +10)^5 (\varphi^4 -4 \varphi^3 +4 \varphi^2 +4 \varphi+1)^2} \\
				&\times\Big[3645\,e^{\frac{2 (8 \varphi^2 +\lambda (4 \varphi +19) \varphi +20)}{2\varphi^2 -7 \varphi +5}} (2 \varphi -5)^5 (-5 \varphi^2 +13 \varphi +10)^2
				\big(420 \varphi^{11} +12 (35 \lambda -426) \varphi^{10} +(1752 \lambda +38015) \varphi^9 \\
				&\: -3 (7756 \lambda +68063) \varphi^8 +3 (41876 \lambda +223849) \varphi^7 -3 (123280 \lambda +442861) \varphi^6 +3 (191480 \lambda +501403) \varphi^5\\
				&\:-15 (13760 \lambda +56677) \varphi^4 -1500 (416 \lambda +15) \varphi^3 +750 (800 \lambda +911) \varphi^2 -555000 \varphi -50000\big) \varphi^8 \\
				& -90\,e^{\frac{2(\varphi^2 +17 \varphi +\lambda (11 \varphi^2 +2 \varphi +10)+10)}{2 \varphi^2 -7 \varphi +5}} (2 \varphi -5) (\varphi^2 -2 \varphi +10)^5\mathcal{F}_0^\alpha(\varphi,\lambda) \\
				& +90\,e^{\frac{2 (7 \varphi^2 +2 \varphi +\lambda (5 \varphi^2 +17 \varphi+10)+10)}{2 \varphi^2 -7 \varphi +5}} (-5 \varphi^2 +13 \varphi +10)^2 \mathcal{F}_1^\alpha(\varphi,\lambda) \\
				& +10\,e^{\frac{38 \varphi +8 \lambda (3 \varphi^2 +5)}{2 \varphi^2 -7 \varphi +5}} (2 \varphi-5) (\varphi ^2-2 \varphi +10 )^5 (5 \varphi^3 -18 \varphi^2 +3 \varphi +10)^2 \mathcal{F}_2^\alpha(\varphi,\lambda) \\
				&+e^{\frac{4 \varphi (3 \varphi+2)+2 \lambda (6 \varphi^2 +15 \varphi +20)}{2 \varphi^2 -7 \varphi +5}} (10-7 \varphi)^2 (2 \varphi^2 -7 \varphi+5) \mathcal{F}_3^\alpha(\varphi,\lambda)\Big]
			\end{split}
		\end{equation}

		\begin{equation}
			\begin{split}
				\mathcal{F}_0^\alpha(\varphi,\lambda)&=	\big(1820 \varphi^{14} -2 (1365 \lambda +283453) \varphi^{13} +(1272048 \lambda +6107021) \varphi^{12}-(20482320 \lambda +20431967) \varphi^{11} \\
				& +(114528546 \lambda +11016989) \varphi^{10} +(60057067-286961148 \lambda ) \varphi^9+516 (600429 \lambda -178658) \varphi^8 \\
				& -6 (5945901 \lambda -1736875) \varphi^7 -3 (57687456 \lambda +1047899) \varphi^6 +(78544260 \lambda +82163557) \varphi^5 \\
				& +5 (3859926 \lambda -8380229) \varphi^4 -25 (224628 \lambda +677189) \varphi^3 -1000 (339 \lambda -2600) \varphi^2 \\
				&+2500 (36 \lambda+307) \varphi -100000\big) \varphi^3
			\end{split}
		\end{equation}

		\begin{equation}
			\begin{split}
				\mathcal{F}_1^\alpha(\varphi,\lambda)&=	\big(56 (756 \lambda^4 +2772 \lambda^3 -5418 \lambda^2 +14286 \lambda -36499) \varphi^{21}
				+4 (5292 \lambda^5 -204876 \lambda^4 -917910 \lambda^3 +1577394 \lambda^2 \\
				&-4683909 \lambda +11923444) \varphi^{20} + (-81648 \lambda^5 +9935568 \lambda^4 +42669684 \lambda^3 -66547764 \lambda^2 +219373890 \lambda \\
				&-580764580) \varphi^{19} -(769500 \lambda^5 +87526440 \lambda^4 +348498828 \lambda^3 -482795622 \lambda^2 +1721592918 \lambda \\
				&-4844351393) \varphi^{18} +3 (5324220 \lambda^5 +196156800 \lambda^4 +710292072 \lambda^3 -871412394 \lambda^2 +3388796130 \lambda \\
				&-9932500151) \varphi^{17} -6 (22655700 \lambda^5 +523413342 \lambda^4 +1691146536 \lambda^3 -1916285607 \lambda^2 +8029146999 \lambda \\
				&-22970685019) \varphi^{16} +6 (130322304 \lambda^5 +2250878958 \lambda^4 +6402258576 \lambda^3 -7151091627 \lambda^2 +31351127040 \lambda \\
				&-80478901804) \varphi^{15} -3 (1125200448 \lambda^5 +15773269572 \lambda^4 +38810417712 \lambda^3 -45760025112 \lambda^2 \\
				&+202797530730 \lambda -429432044573) \varphi^{14} +3 (3776764320 \lambda^5 +45175721580 \lambda^4 +94679143620 \lambda^3 \\
				&-124446510828 \lambda^2 +538795811808 \lambda-874763497805) \varphi^{13} - (30031456320 \lambda^5 +315862563960 \lambda^4 \\
				& +561487253904 \lambda^3 -838042037298 \lambda^2 +3455249200086 \lambda -4062109868821) \varphi^{12} \\
				&+ (62201606400 \lambda^5 +593632671480 \lambda^4 +916740799560 \lambda^3 -1494677901474 \lambda^2 +5754577240896 \lambda \\
				& -4668350180471) \varphi^{11} -5 (19580356800 \lambda^5 +176536113120 \lambda^4 +260955497808 \lambda^3 -410059014300 \lambda^2 \\
				& +1425250272948 \lambda -738352287991) \varphi^{10} + (109657800000 \lambda^5 +1008438606000 \lambda^4 +1736574508800 \lambda^3 \\
				& -2093664647520 \lambda^2 +6044598866340 \lambda -1451240592997) \varphi^9 -30 (2406600000 \lambda^5 +28367730000 \lambda^4 \\
				& +70338417600 \lambda^3 -45328948680 \lambda^2 +89583315096 \lambda +22706537389) \varphi^8 +300 (25920000 \lambda^5 \\
				&+1608435000 \lambda^4 +6720213000 \lambda^3 +246911220 \lambda^2 -2375573988 \lambda +4961375143) \varphi^7 +3000 (7560000 \lambda^5 \\
				&-40113000 \lambda^4 -421821000 \lambda^3 -425184150 \lambda^2 +607314348 \lambda -322995619) \varphi^6 -7500 (1440000 \lambda^5 \\
				& +5796000 \lambda^4 -54639600 \lambda^3 -153034500 \lambda^2 +99662580 \lambda -21029827) \varphi^5 +600000 (49500 \lambda^4 \\
				& -17400 \lambda^3 -553815 \lambda^2 -294795 \lambda +131716) \varphi^4	-250000 (104400 \lambda^3 +32580 \lambda^2 -418236 \lambda -83899) \varphi^3 \\
				&+10000000 (1170 \lambda^2 +1482 \lambda -1517) \varphi^2 -12500000 (276 \lambda +227) \varphi +500000000\big) \varphi^3
			\end{split}
		\end{equation}

		\begin{equation}
			\begin{split}
				\mathcal{F}_2^\alpha(\varphi,\lambda)&= 91 (6 \lambda -1) \varphi^{11}-3 (364 \lambda^2 +55180 \lambda -9227) \varphi^{10} +(495060 \lambda^2 +392514 \lambda -93009) \varphi^9 \\
				&-3 (637528 \lambda^2 +57808 \lambda -31441) \varphi^8 +12 (198731 \lambda^2 +28532 \lambda -7164) \varphi^7 \\
				& -12 (64348 \lambda^2 +85930 \lambda -19077) \varphi^6 -3 (98092 \lambda^2 -152474 \lambda +82699) \varphi^5	+3 (23324 \lambda^2 +87008 \lambda +10911) \varphi^4 \\
				& +3 (1760 \lambda^2 -5386 \lambda +15399) \varphi^3 +(-1200 \lambda^2 -16920 \lambda +127) \varphi^2 +120 (15 \lambda -26) \varphi +300
			\end{split}
		\end{equation}

		\begin{equation}
			\begin{split}
				\mathcal{F}_3^\alpha(\varphi,\lambda)&= 50 (3240 \lambda^6 -1296 \lambda^5 -88290 \lambda^4 +381960 \lambda^3 -652950 \lambda^2 +727205) \varphi^{24} \\
				&-20 (147420 \lambda^6 -44928 \lambda^5 -4204170 \lambda^4 +18602280 \lambda^3 -33120150 \lambda^2 +38129261) \varphi^{23} \\
				&+6 (5261220 \lambda^6 -1254528 \lambda^5 -152957520 \lambda^4 +681347280 \lambda^3 -1235032800 \lambda^2 +1429039213)\varphi^{22} \\
				&-4 (59489640 \lambda^6 -10101024 \lambda^5 -1753218270 \lambda^4 +7831194840 \lambda^3 -14395203090 \lambda^2 +16633125349) \varphi^{21} \\
				&+10 (137458944 \lambda^6 -13513392 \lambda^5 -4058563968 \lambda^4 +17982345816 \lambda^3 -33108357720 \lambda^2 +37141486681)\varphi^{20} \\
				&-108 (58733040 \lambda^6 -1173168 \lambda^5 -1717020480 \lambda^4 +7435001560 \lambda^3 -13412523590 \lambda^2 +13716168573) \varphi^{19} \\
				&+3 (7927494840 \lambda^6 +566984448 \lambda^5 -226440651600 \lambda^4 +942045890400 \lambda^3 -1623609048240 \lambda^2 \\
				&+1377919510577) \varphi^{18} -120 (609357762 \lambda^6 +110306232 \lambda^5 -16705018989 \lambda^4 +65225560752 \lambda^3 \\
				&-104174444055 \lambda^2 +63849080257) \varphi^{17} +36 (5089341420 \lambda^6 +1643165568 \lambda^5 -131051545065 \lambda^4 \\
				&+464821578840 \lambda^3 -660083367795 \lambda^2 +219178925692) \varphi^{16} -10 (36793942848 \lambda^6 +19084009536 \lambda^5 \\
				&-859901715036 \lambda^4 +2620890216720 \lambda^3 -3090418835040 \lambda^2 -23109700363) \varphi^{15} +4 (142030013760 \lambda^6 \\
				& +117441007488 \lambda^5 -2837610484470 \lambda^4+6479819264520 \lambda^3 -5177085066420 \lambda^2 -3342422380003) \varphi^{14} \\
				&-48 (12619876680 \lambda^6 +18316440576 \lambda^5-182340360450 \lambda^4 +114920829840 \lambda^3 +196010948685 \lambda^2 \\
				&-363835575266) \varphi^{13} + (294722582400 \lambda^6 +1177249946112 \lambda^5 +847017479520 \lambda^4 -26649573192000 \lambda^3 \\
				&+36203253733800 \lambda^2 -4562867298487) \varphi^{12} +10 (30754598400 \lambda^6 -95322396672 \lambda^5 -1085731483104 \lambda^4 \\
				&+3937302440640 \lambda^3 -2986696472016 \lambda^2 -1205877126071) \varphi^{11} -90 (7157880000 \lambda^6 +405319680 \lambda^5 \\
				&-115067194560 \lambda^4 +153423041856 \lambda^3 +40325100288 \lambda^2 -148742987459) \varphi^{10} +2160 (162150000 \lambda^6 \\
				&+457416000 \lambda^5 -120261600 \lambda^4 -9008824160 \lambda^3 +10384187828 \lambda^2 -470544233) \varphi^9 \\
				&+120 (2384100000 \lambda^6 -8406720000 \lambda^5 -55729485000 \lambda^4 +172695967200 \lambda^3 -74595745740 \lambda^2 \\
				&-60960277757) \varphi^8 -2880 (76500000 \lambda^6 +77220000 \lambda^5 -410343750 \lambda^4 -2276760000 \lambda^3 \\
				&+3642388350 \lambda^2 -1645337107) \varphi^7 -1600 (60750000 \lambda^6 -304560000 \lambda^5 -784788750 \lambda^4 \\
				&+2722590000 \lambda^3 -90205875 \lambda^2 -502947169) \varphi^6 +16000 (7560000 \lambda^5 -12420000 \lambda^4 \\
				&-64818000 \lambda^3 +82405875 \lambda^2 +27272024) \varphi^5 -300000 (234000 \lambda^4 -153600 \lambda^3 -1111020 \lambda^2 -438251) \varphi^4 \\
				&+1000000 (28800 \lambda^3 -2160 \lambda^2 -27067) \varphi^3 -1000000 (6600 \lambda^2 +14123) \varphi^2 -720000000 \varphi +300000000
			\end{split}
		\end{equation}

		\end{widetext}

		\item Isochoric compressibility $\bar\beta_V=\bar\beta_V^{CS}+\Delta\bar\beta_V $

		Its expression depends on a remaining integral term that is not known explicitly, defined as follows:
		\begin{equation}
			\mathcal{V}(\varphi,\lambda)=-\int_0^\varphi\frac{(1+2\psi)^4}{(1-\psi)^8}\,\Delta S\,d\psi\,,
		\end{equation}
		where $\Delta S$ has been defined in Eq.~(\ref{eqS0HT}).

		\begin{widetext}
		\begin{equation}
			\begin{split}
				\Delta\bar\beta_V &=\frac{9720 (\varphi -1)^3 \varphi^3 (35 \varphi^3 -141 \varphi^2 +60 \varphi +100)(\varphi^2 -2\varphi+10)^4 \mathcal{V}(\varphi,\lambda)}
				{14580 (T^*)^2 (10-7 \varphi) \varphi^4 (-5 \varphi^2 +13 \varphi +10) (\varphi^2 -2 \varphi+10)^4 (-\varphi^3 +\varphi^2 +\varphi +1)} \\
				&+\frac{e^{\frac{2 (3 \varphi(2 \varphi-5) +\lambda (6 \varphi^2 +13 \varphi +20))}{(1-\varphi)(2 \varphi-5)}}}{14580 (T^*)^2 (10-7 \varphi) \varphi^4 (-5 \varphi^2 +13 \varphi +10)
				(\varphi^2 -2 \varphi+10)^4 (-\varphi^3 +\varphi^2 +\varphi+1)} \\
				&\times\Big[10935\,e^{\frac{8 (\lambda +2) \varphi^2 +(34 \lambda -38) \varphi +40}{2 \varphi^2 -7 \varphi +5}} (5-2\varphi)^6 (25 \varphi^5 -125 \varphi^4 +151 \varphi^3 +128 \varphi^2 -415 \varphi -250) \varphi^9 \\
				&-90\,e^{\frac{2 (\varphi^2 -2 \varphi +\lambda (11 \varphi^2 +10)+10)}{2 \varphi^2 -7 \varphi+5}} (5-2 \varphi)^2 (\varphi^2 -2 \varphi +10)^4
				(13 \varphi^7 -5875 \varphi^6 +14376 \varphi^5 -7853 \varphi^4 -2026 \varphi^3 \\
				&\:+609 \varphi^2 +37 \varphi -10) \varphi^3
				+90\,e^{\frac{2 (7 \varphi^2 -17 \varphi +5 \lambda (\varphi^2 +3 \varphi +2)+10)}{2 \varphi^2 -7 \varphi +5}} (10 \varphi^3 -51 \varphi^2 +45 \varphi +50)\mathcal{F}_0^\beta(\varphi,\lambda) \\
				&+10\,e^{\frac{4 \lambda (6 \varphi^2 -\varphi +10)}{(\varphi -1)(2 \varphi -5)}} (\varphi-1)^2 ((6 \lambda -1) \varphi +1) (\varphi^2 -2 \varphi +10)^4 
				(65 \varphi^9 -29544 \varphi^8 +148125 \varphi^7 -167403 \varphi^6 \\
				&\:-51801 \varphi^5 +107913 \varphi^4 +12528 \varphi^3 -6621 \varphi^2 -240 \varphi+100) \\
				&+e^{\frac{2 (3 \varphi(2 \varphi -5) +\lambda (6 \varphi^2 +13 \varphi +20))}{(\varphi -1)(2 \varphi -5)}}(7 \varphi^2 -17 \varphi +10) \mathcal{F}_1^\beta(\varphi,\lambda)\Big]
			\end{split}
		\end{equation}

		\begin{equation}
			\begin{split}
				\mathcal{F}_0^\beta(\varphi,\lambda)&=\big(2 (756 \lambda^4 +2772 \lambda^3 -5418 \lambda^2 +14286 \lambda -36499) \varphi^{14} \\
				& +(-15768 \lambda^4 -70164 \lambda^3 +85158 \lambda^2 -338730 \lambda +995995) \varphi^{13} \\
				&+2 (55647 \lambda^4 +237240 \lambda^3 -298998 \lambda^2 +1227303 \lambda -3067243) \varphi^{12} \\
				&-2 (260091 \lambda^4 +1101330 \lambda^3 -1548819 \lambda^2 +5822409 \lambda -10810537) \varphi^{11} \\
				&+(1807596 \lambda^4 +7232220 \lambda^3 -11541366 \lambda^2 +38476608 \lambda -46772437) \varphi^{10} \\
				&-2 (2333286 \lambda^4 +8754534 \lambda^3 -15276033 \lambda^2 +43989534 \lambda -29969155) \varphi^9 \\
				& +3 (2859840 \lambda^4 +10314576 \lambda^3 -16553232 \lambda^2 +41992428 \lambda -10702087) \varphi^8 \\
				&-6 (1800900 \lambda^4 +6317280 \lambda^3 -4445091 \lambda^2 +12407964 \lambda +5158313) \varphi^7 \\
				& +(6318000 \lambda^4 +33771600 \lambda^3 +58639140 \lambda^2 -68619852 \lambda +77996247) \varphi^6 \\
				& +4 (472500 \lambda^4 -5395500 \lambda^3 -27422325 \lambda^2 +25824630 \lambda -11532547) \varphi^5 \\
				&-10 (270000 \lambda^4 -504000 \lambda^3 -6258600 \lambda^2 +37740 \lambda +887557) \varphi^4 \\
				&+100 (18000 \lambda^3 -124200 \lambda^2 -178020 \lambda +11459) \varphi^3-1000 (900 \lambda^2 -240 \lambda -3007) \varphi^2 \\
				&+10000 (30 \lambda +1) \varphi -50000\big) \varphi^3
			\end{split}
		\end{equation}

		\begin{equation}
			\begin{split}
				\mathcal{F}_1^\beta(\varphi,\lambda)&= 10 (3240 \lambda^6 -1296 \lambda^5 -88290 \lambda^4 +381960 \lambda^3 -652950 \lambda^2 +727205) \varphi^{18} \\
				& -12 (29970 \lambda^6 -4968 \lambda^5 -910170 \lambda^4 +4144680 \lambda^3 -7745250 \lambda^2 +9247363) \varphi^{17} \\
				& +12 (236520 \lambda^6 -14256 \lambda^5 -6905925 \lambda^4 +30342060 \lambda^3 -55886715 \lambda^2 +61704296) \varphi^{16} \\
				& -30 (495180 \lambda^6 +59184 \lambda^5 -14231160 \lambda^4 +59329128 \lambda^3 -103502454 \lambda^2 +91484027) \varphi^{15} \\
				& +12 (4981500 \lambda^6 +1453032 \lambda^5 -134414775 \lambda^4 +515280060 \lambda^3 -821818950 \lambda^2 +475049483) \varphi^{14} \\
				& -12 (15260400 \lambda^6 +8075376 \lambda^5 -380042955 \lambda^4 +1280936160 \lambda^3 -1778104530 \lambda^2 +385117384) \varphi^{13} \\
				& +9 (46638720 \lambda^6 +41043456 \lambda^5 -1034475840 \lambda^4 +2839937120 \lambda^3 -3144996700 \lambda^2 -767414941) \varphi^{12} \\
				& -12 (60283440 \lambda^6 +83115072 \lambda^5 -1075705920 \lambda^4 +1843165800 \lambda^3 -1068168315 \lambda^2 -1846059662) \varphi^{11} \\
				& +3 (237168000 \lambda^6 +690571008 \lambda^5 -2712547440 \lambda^4 -3780691200 \lambda^3 +8800066260 \lambda^2 -6471065803) \varphi^{10} \\
				& -5 (60912000 \lambda^6 +523376640 \lambda^5 +1102206528 \lambda^4 -10121178816 \lambda^3 +9189146232 \lambda^2 +1353711563) \varphi^9 \\
				& -120 (6480000 \lambda^6 -18230400 \lambda^5 -141056100 \lambda^4 +390936168 \lambda^3 -127383837 \lambda^2 -231030331) \varphi^8 \\
				& +60 (8100000 \lambda^6 +16848000 \lambda^5 -24408000 \lambda^4 -393811200 \lambda^3 +545066784 \lambda^2 -376608625) \varphi^7 \\
				& +120 (2700000 \lambda^6 -14040000 \lambda^5 -38340000 \lambda^4 +136884000 \lambda^3 -8082600 \lambda^2 +31390829) \varphi^6 \\
				& -480 (540000 \lambda^5 -1856250 \lambda^4 -3735000 \lambda^3 +9140250 \lambda^2 +1949411) \varphi^5 \\
				& +12000 (13500 \lambda^4 -27000 \lambda^3 -84600 \lambda^2 -58433) \varphi^4 -360000 (200\lambda^3 -175 \lambda^2 -354) \varphi^3 \\
				& +1200000 (15 \lambda^2 +52) \varphi^2 -1500000 \varphi -1000000
			\end{split}
		\end{equation}

		\end{widetext}

		\item Heat Capacities $\bar C_P-\bar C_V = (\bar C_P-\bar C_V)^{CS}+\Delta\bar C_{PV}$

		\begin{widetext}
		\begin{equation}
			\begin{split}
				\Delta\bar C_{PV}&=-\frac{e^{-\frac{4 (3 \varphi (\varphi +2)+\lambda (7 \varphi^2 +11 \varphi +30))}{(\varphi -1) (2 \varphi -5)}}(-\varphi^3 +\varphi^2 +\varphi +1)^2}
				{4860 T^* (10-7 \varphi)^2 (\varphi -1)^3 \varphi^4 (2 \varphi-5)(-5 \varphi^2 +13 \varphi +10)^2 (\varphi^2 -2 \varphi +10)^5 (\varphi^4 -4 \varphi^3 +4 \varphi^2 +4 \varphi +1)^2} \\
				&\times\Big[3645\,e^{\frac{4 (4 \varphi^2 +4 \varphi +\lambda (6 \varphi^2 +13 \varphi +20)+10)}{(\varphi -1)(2 \varphi -5)}} (2 \varphi -5)^5 (-5 \varphi^2 +13 \varphi +10)^2
				(420\varphi^{11}+12 (35 \lambda -426) \varphi^{10} \\
				&\:+(1752 \lambda +38015) \varphi^9 -3 (7756 \lambda +68063) \varphi^8 +3 (41876 \lambda+223849) \varphi^7 -3 (123280 \lambda +442861) \varphi^6 \\
				&\:+3 (191480 \lambda +501403) \varphi^5 -15 (13760 \lambda +56677) \varphi^4 -1500 (416 \lambda +15) \varphi^3 +750 (800 \lambda +911) \varphi^2 \\
				&\:-555000 \varphi -50000) \varphi ^8
				-90\,e^{\frac{2 (\varphi^2 +25 \varphi +\lambda (19 \varphi^2 +9 \varphi +50)+10)}{(\varphi -1)(2 \varphi -5)}} (2 \varphi -5)(\varphi^2 -2 \varphi +10)^5(1820 \varphi^{14} \\
				&\:-2 (1365 \lambda +283453) \varphi ^{13}+(1272048 \lambda +6107021) \varphi^{12} -(20482320 \lambda +20431967) \varphi^{11} \\
				&+(114528546 \lambda +11016989) \varphi^{10} +(60057067-286961148 \lambda) \varphi^9+516 (600429 \lambda -178658) \varphi^8 \\
				&\:-6 (5945901 \lambda -1736875) \varphi^7 -3 (57687456 \lambda +1047899) \varphi^6 +(78544260 \lambda +82163557) \varphi^5 \\
				&\:+5 (3859926 \lambda -8380229) \varphi^4 -25 (224628 \lambda +677189) \varphi^3 -1000 (339\lambda -2600) \varphi^2 \\
				&\:+2500 (36 \lambda +307) \varphi-100000) \varphi^3
				+90\,e^{\frac{2 (7 \varphi^2 +10 \varphi +\lambda (13 \varphi^2 +24 \varphi +50)+10)}{2 \varphi^2 -7 \varphi +5}} (-5 \varphi^2 +13 \varphi +10)^2\mathcal{F}^C_0(\varphi,\lambda) \\
				&+10\,e^{\frac{54 \varphi +2 \lambda (20 \varphi^2 +7 \varphi +60)}{2 \varphi^2 -7 \varphi +5}} (2 \varphi -5) (\varphi^2 -2 \varphi +10)^5
				(5 \varphi^3-18 \varphi^2+3 \varphi +10)^2\mathcal{F}_1^C(\varphi,\lambda) \\
				&+e^{\frac{4 (3 \varphi (\varphi +2)+\lambda (7 \varphi^2 +11 \varphi +30))}{(\varphi -1)(2 \varphi -5)}} (10-7 \varphi)^2 (2 \varphi^2 -7 \varphi +5)\mathcal{F}_2^C(\varphi,\lambda)\Big]
			\end{split}
		\end{equation}

		\begin{equation}
			\begin{split}
				\mathcal{F}_0^C(\varphi,\lambda)&= \big(56 (756 \lambda^4 +2772 \lambda^3 -5418 \lambda^2 +14286 \lambda -36499) \varphi^{21} \\
				&+4 (5292 \lambda^5-204876 \lambda^4 -917910 \lambda^3 +1577394 \lambda^2 -4683909 \lambda +11923444) \varphi^{20} \\
				&+(-81648 \lambda^5+9935568 \lambda^4 +42669684 \lambda^3 -66547764 \lambda^2 +219373890 \lambda -580764580) \varphi^{19} \\
				&-(769500 \lambda^5 +87526440 \lambda^4 +348498828 \lambda^3 -482795622 \lambda^2 +1721592918 \lambda -4844351393) \varphi^{18} \\
				&+3 (5324220 \lambda^5 +196156800 \lambda^4 +710292072 \lambda^3 -871412394 \lambda^2 +3388796130 \lambda -9932500151) \varphi^{17} \\
				&-6 (22655700 \lambda^5 +523413342 \lambda^4 +1691146536 \lambda^3 -1916285607 \lambda^2 +8029146999 \lambda -22970685019) \varphi^{16} \\
				&+6 (130322304 \lambda^5 +2250878958 \lambda^4 +6402258576 \lambda^3 -7151091627 \lambda^2 +31351127040 \lambda -80478901804) \varphi^{15} \\
				&-3 (1125200448 \lambda^5 +15773269572 \lambda^4 +38810417712 \lambda^3 -45760025112 \lambda^2+202797530730 \lambda \\
				&\: -429432044573) \varphi^{14} +3 (3776764320 \lambda^5 +45175721580 \lambda^4 +94679143620 \lambda^3-124446510828 \lambda^2 \\
				&\:+538795811808 \lambda -874763497805) \varphi^{13} -(30031456320 \lambda^5 +315862563960 \lambda^4 +561487253904 \lambda^3 \\
				&\: -838042037298 \lambda^2 +3455249200086 \lambda -4062109868821) \varphi^{12} +(62201606400 \lambda^5 +593632671480 \lambda^4 \\
				&\:+916740799560 \lambda^3 -1494677901474 \lambda^2+5754577240896 \lambda -4668350180471) \varphi^{11} -5 (19580356800 \lambda^5 \\
				&\:+176536113120 \lambda^4 +260955497808 \lambda^3 -410059014300 \lambda^2 +1425250272948 \lambda -738352287991) \varphi^{10} \\
				&+(109657800000 \lambda^5 +1008438606000 \lambda^4 +1736574508800 \lambda^3 -2093664647520 \lambda^2 +6044598866340 \lambda \\
				&\:-1451240592997) \varphi^9 -30 (2406600000 \lambda^5 +28367730000 \lambda^4 +70338417600 \lambda^3 -45328948680 \lambda^2 \\
				&\:+89583315096 \lambda +22706537389) \varphi^8 +300 (25920000 \lambda^5 +1608435000 \lambda^4 +6720213000 \lambda^3 \\
				&\:+246911220 \lambda^2 -2375573988 \lambda +4961375143) \varphi^7 +3000 (7560000 \lambda^5 -40113000 \lambda^4 \\
				&\:-421821000 \lambda^3 -425184150 \lambda^2 +607314348 \lambda -322995619) \varphi^6 -7500 (1440000 \lambda^5 \\
				&\:+5796000 \lambda^4 -54639600 \lambda^3 -153034500 \lambda^2 +99662580 \lambda -21029827) \varphi^5+600000 (49500 \lambda^4 -17400 \lambda^3 \\
				&\:-553815 \lambda^2 -294795 \lambda +131716) \varphi^4 -250000 (104400 \lambda^3 +32580 \lambda^2 -418236 \lambda-83899) \varphi^3 \\
				& +10000000 (1170 \lambda^2 +1482 \lambda -1517) \varphi^2 -12500000 (276 \lambda +227) \varphi +500000000\big) \varphi^3
			\end{split}
		\end{equation}

		\begin{equation}
			\begin{split}
				\mathcal{F}_1^C(\varphi,\lambda)&= 91 (6 \lambda -1)\varphi^{11} -3 (364 \lambda^2 +55180 \lambda -9227) \varphi^{10} +(495060 \lambda^2 +392514 \lambda -93009) \varphi^9 \\
				&-3 (637528 \lambda^2 +57808 \lambda -31441) \varphi^8 +12 (198731 \lambda^2 +28532 \lambda -7164)\varphi^7 -12 (64348 \lambda^2 +85930 \lambda -19077) \varphi^6 \\
				& -3 (98092 \lambda^2 -152474 \lambda +82699)\varphi^5 +3 (23324 \lambda^2 +87008 \lambda +10911) \varphi^4 +3 (1760 \lambda^2 -5386 \lambda +15399) \varphi^3 \\
				&+(-1200 \lambda^2 -16920 \lambda +127) \varphi^2 +120 (15 \lambda -26) \varphi +300
			\end{split}
		\end{equation}

		\begin{equation}
			\begin{split}
				\mathcal{F}_2^C(\varphi,\lambda)&= 50 (3240 \lambda^6 -1296 \lambda^5 -88290 \lambda^4 +381960 \lambda^3 -652950 \lambda^2 +727205) \varphi^{24} \\
				& -20 (147420 \lambda^6 -44928 \lambda^5 -4204170 \lambda^4 +18602280 \lambda^3 -33120150 \lambda^2 +38129261) \varphi^{23} \\
				& +6 (5261220 \lambda^6 -1254528 \lambda^5 -152957520 \lambda^4 +681347280 \lambda^3 -1235032800 \lambda^2 +1429039213) \varphi^{22} \\
				&-4 (59489640 \lambda^6 -10101024 \lambda^5 -1753218270 \lambda^4 +7831194840 \lambda^3 -14395203090 \lambda^2 +16633125349) \varphi^{21} \\
				&+10 (137458944 \lambda^6 -13513392 \lambda^5 -4058563968 \lambda^4 +17982345816 \lambda^3 -33108357720 \lambda^2 +37141486681) \varphi^{20} \\
				& -108 (58733040 \lambda^6 -1173168 \lambda^5 -1717020480 \lambda^4 +7435001560 \lambda^3 -13412523590 \lambda^2 +13716168573) \varphi^{19} \\
				&+3 (7927494840 \lambda^6 +566984448 \lambda^5 -226440651600 \lambda^4 +942045890400 \lambda^3 -1623609048240 \lambda^2 \\
				&\:+1377919510577) \varphi^{18} -120 (609357762 \lambda^6 +110306232 \lambda^5 -16705018989 \lambda^4 +65225560752 \lambda^3 \\
				&-104174444055 \lambda^2 +63849080257) \varphi^{17} +36 (5089341420 \lambda^6 +1643165568 \lambda^5 -131051545065 \lambda^4 \\
				&\:+464821578840 \lambda^3 -660083367795 \lambda^2 +219178925692) \varphi^{16} -10 (36793942848 \lambda^6 +19084009536 \lambda^5 \\
				&\:-859901715036 \lambda^4 +2620890216720 \lambda^3 -3090418835040 \lambda^2 -23109700363) \varphi^{15} +4 (142030013760 \lambda^6 \\
				&\:+117441007488 \lambda^5 -2837610484470 \lambda^4 +6479819264520 \lambda^3 -5177085066420 \lambda^2 -3342422380003) \varphi^{14} \\
				&-48 (12619876680 \lambda^6 +18316440576 \lambda^5 -182340360450 \lambda^4 +114920829840 \lambda^3 +196010948685 \lambda^2 \\
				&\:-363835575266) \varphi^{13} +(294722582400 \lambda^6 +1177249946112 \lambda^5 +847017479520 \lambda^4 -26649573192000 \lambda^3 \\
				&\:+36203253733800 \lambda^2 -4562867298487) \varphi^{12} +10 (30754598400 \lambda^6 -95322396672 \lambda^5 -1085731483104 \lambda^4 \\
				&\:+3937302440640 \lambda^3 -2986696472016 \lambda^2 -1205877126071) \varphi^{11} -90 (7157880000 \lambda^6 +405319680 \lambda^5 \\
				&\:-115067194560 \lambda^4 +153423041856 \lambda^3 +40325100288 \lambda^2 -148742987459) \varphi^{10} +2160 (162150000 \lambda^6 \\
				&+457416000 \lambda^5 -120261600 \lambda^4 -9008824160 \lambda^3 +10384187828 \lambda^2 -470544233) \varphi^9 \\
				&+120 (2384100000 \lambda^6 -8406720000 \lambda^5 -55729485000 \lambda^4 +172695967200 \lambda^3 -74595745740 \lambda^2 \\
				&\:-60960277757) \varphi^8 -2880 (76500000 \lambda^6 +77220000 \lambda^5 -410343750 \lambda^4 -2276760000 \lambda^3 \\
				&\:+3642388350 \lambda^2 -1645337107) \varphi^7 -1600 (60750000 \lambda^6 -304560000 \lambda^5 -784788750 \lambda^4 \\
				&+2722590000 \lambda^3 -90205875 \lambda^2 -502947169) \varphi^6 +16000 (7560000 \lambda^5 -12420000\lambda^4 -64818000 \lambda^3 \\
				&\:+82405875 \lambda^2 +27272024) \varphi^5 -300000 (234000 \lambda^4 -153600 \lambda^3 -1111020\lambda^2 -438251) \varphi^4 \\
				&+1000000 (28800 \lambda^3-2160 \lambda^2-27067) \varphi^3-1000000 (6600 \lambda^2 +14123) \varphi^2 -720000000 \varphi +300000000
			\end{split}
		\end{equation}

		\end{widetext}

	\end{itemize}

\bibliography{Struct2.bib}

\end{document}